\documentclass[9pt,shortpaper,twoside,web]{ieeecolor}
\usepackage{generic}
\usepackage{cite}
\usepackage{amsmath,amssymb,amsfonts}
\usepackage{algorithmic}
\usepackage{graphicx}
\usepackage{textcomp}
\usepackage{url}
\usepackage{xcolor}

\usepackage{amsthm}
\usepackage{cancel}
\def\BibTeX{{\rm B\kern-.05em{\sc i\kern-.025em b}\kern-.08em
    T\kern-.1667em\lower.7ex\hbox{E}\kern-.125emX}}
\theoremstyle{plain}
\newtheorem{theorem}{Theorem}
\newtheorem{lemma}{Lemma}
\newtheorem{corollary}{Corollary}

\theoremstyle{assumption}
\newtheorem{Hypothesis}{Hypothesis}

\theoremstyle{assumption}
\newtheorem{definition}{Definition}

\newtheorem{remark}{Remark}
\theoremstyle{goal}

\theoremstyle{assumption}
\newtheorem{assumption}{Assumption}

\begin{document}
\title{Exponentially Stable Adaptive Observation {for Systems Parameterized by Unknown Physical Parameters}}
\author{Anton Glushchenko, \IEEEmembership{Member, IEEE}, Konstantin Lastochkin
\thanks{Research was in part financially supported by Grants Council of the President of the Russian Federation (project MD-1787.2022.4).}
\thanks{A. I. Glushchenko is with V.A. Trapeznikov Institute of Control Sciences RAS, Moscow, Russia (phone: +79102266946; e-mail: aiglush@ipu.ru).}
\thanks{K. A. Lastochkin is with V.A. Trapeznikov Institute of Control Sciences RAS, Moscow, Russia (e-mail: lastconst@ipu.ru).}}

\maketitle

\begin{abstract}
The method to design exponentially stable adaptive observers is proposed for linear time-invariant systems parameterized by unknown {physical} parameters. Unlike existing adaptive solutions, the system state-space matrices {\it{\textbf{A, B}}} are not restricted to be represented in the observer canonical form to implement the observer. The original system description is used instead, and, consequently, the original state vector is obtained. The class of systems for which the method is applicable is identified via three assumptions related to: \linebreak {({\it{\textbf{i}}}) the boundedness of a control signal and all system trajectories, ({\it{\textbf{ii}}}) the identifiability of the {physical} parameters of {\it{\textbf{A}}} and {\it{\textbf{B}}} from the numerator and denominator polynomials of a system input/output transfer function and ({\it{\textbf{iii}}}) the complete observability of system states}. In case they are met and the regressor is finitely exciting, the proposed adaptive observer, which is based on the known GPEBO and DREM procedures, ensures exponential convergence of both system parameters and states estimates to their true values. Detailed analysis for stability and convergence has been provided along with simulation results to validate the developed theory.
\end{abstract}


\section{Introduction}
\label{sec:introduction}

Installation of a complete set of sensors is usually impracticable due to their high cost and vulnerability to adverse effect of the environment: vibration, humidity, dust conditions, etc. Therefore, in order to solve the fault detection and control problems effectively, the unmeasurable states of the system are often reconstructed with the help of various observers. When the system parameters are uncertain, then an actual problem is to combine state observation with parameter estimation \cite{b1}. Without pretending to provide an exhaustive review, the main methods of adaptive observers design are considered below.

{In the 1970s several different approaches to design adaptive observers were proposed in the control literature \cite{b2, b3, b4, b5, b6, b7}. Carroll and Lindorff \cite{b2} were the first who developed a method of adaptive observation of both unmeasured state and unknown parameters simultaneously. Luders and Narendra suggested an alternative observer \cite{b3} and later modified it to have a simpler structure \cite{b4}. The paper by Kudva and Narendra \cite{b5} proposed yet another method, and in \cite{b4, b6} Narendra and Kudva showed that all these results \cite{b2, b3, b5} could be derived in a unified manner. Kreisselmeier \cite{b7} proposed a parametrization that, unlike \cite{b2, b3, b5}, allows one to: {\it{i}}) transform the adaptive observer design problem into the one of estimation of a linear regression equation (LRE) unknown parameters; {\it{ii}}) separate completely the observer dynamics from the adaptive loop to make the design of suitable parameter adaptation schemes substantially simpler. All above-considered designs provide exponential convergence of the system states estimates to their true values when the strict persistent excitation (PE) requirement is met.}

In \cite{b8, b9, b10, b11} various adaptive observers have been proposed for SISO and MIMO systems, which provide exponentially stable observation when strictly weaker finite excitation condition (FE) is satisfied. The disadvantages of the solutions \cite{b8, b9, b10, b11} are the high dynamic order of the parametrizations-in-use, the heuristics used to improve the properties of the estimates (particularly, multiple switches in adaptive law at predefined time instances), and the insensitivity of the identification laws to changes in the system parameters. In \cite{b12} an observer is proposed that ensures exponential observation under lack of excitation of any order greater than zero. The disadvantage of this solution is the need to compute the eigen-decomposition of the extended regressor matrix online. To relax the requirement of the regressor persistent excitation, an adaptive observer with an identifier based on the ridge regression technique is proposed in \cite{b13}. The properties of this solution are demonstrated experimentally but have not been proved analytically.  A more elegant relaxation of the PE requirement for the adaptive observation problem has been proposed using the parameter estimation-based observer (PEBO) design \cite{b14}. In contrast to \cite{b2, b3, b4, b5, b6, b8, b9, b11, b12, b13}, state estimates are obtained by means of an algebraic equation that relates system states with the product of known signals (regressors) by the unknown parameters. Estimates of these parameters are calculated using a finite-time identification technique \cite{b15} that requires the regressor to be FE. Such estimator is derived from a measurable regression equation obtained using the dynamic regressor extension and mixing (DREM) procedure \cite{b16}.

{Regardless of convergence conditions, the above-considered solutions \cite{b2, b3, b4, b5, b6, b7, b8, b9, b10, b11, b12, b13, b14} ultimately demand that the system is represented in the observer canonical form. However, states $\xi \left( t \right)$ of such form are useless to solve a number of practical control problems because they are virtual and related to the plant original states $x\left( t \right)$ via the linear transformation $\xi \left( t \right) = Tx\left( t \right)$.} 

{To the best of authors’ knowledge, all existing observers to reconstruct original system states instead of virtual ones may be devoted into two main groups.}

{Approaches from the first group exploit such system properties as strict passivity or triangular structure. In \cite{b17} an observer is proposed that allows one to reconstruct original states if the Strict Positive Realness (SPR) condition is met. In contrast to \cite{b2, b3, b4, b5, b6, b7, b8, b9, b10, b11, b12, b13, b14}, the method does not require the regressor to be persistently or finitely exciting for asymptotic convergence of the states observation error. In \cite{b18}, the SPR requirement is relaxed with a high gain observer to obtain an auxiliary signal, which has relative degree one with respect to the unknown parameters. In \cite{b19} for nonlinear systems of triangular structure an observer is proposed to reconstruct the original states. In \cite{b20}, a hybrid adaptive observer was developed for nonlinear systems represented in the Brunovsky form. However, the class of completely observable systems is wider than the systems with special properties from these studies, which is their main drawback.}

{Approaches from the second group \cite{b2, b21, b22} try to implement the dynamic inversion-based change of coordinates to obtain the original states estimates from virtual ones (e.g. $\hat x\left( t \right) = {\hat T_I}\left( t \right)\hat \xi \left( t \right)$, where ${T_I}{\rm{: = }}{T^{ - 1}}$). But the only available information to do this is estimates of virtual states $\hat \xi \left( t \right)$ and parameters ${\hat \psi _{ab}}\left( t \right)$ of the system input/output transfer function, that leads to two problems: 1) identifiability of the inverse transform ${T_I}$ from the parameters ${\psi _{ab}}$ of system transfer function (i.e. input/output data), 2) the estimates ${\hat T_I}\left( t \right)$ are to be obtained under the condition that the system parameters are unknown.}

{The first problem has a positive solution for identifiable in the sense of \cite{b23, b24, b25} linear time-invariant systems overparameterized by unknown physical parameters $\theta$. Considering the identifiable systems, the transfer function parameters ${\psi _{ab}}\left( \theta  \right)$ could be recalculated into physical parameters $\theta$ with the help of some function ${\cal F}$. Therefore, as the inverse transform ${T_I}$ is also parameterized by $\theta$, then ${T_I}\left( \theta  \right){\rm{: = }}{T_I}$ could be calculated via composition ${T_I}\left( \theta  \right) = \left( {{T_I} \circ {\cal F}} \right)\left( {{\psi _{ab}}} \right)$.}

{Straightforward solution \cite{b2, b21} of the second problem is a certainty equivalence substitution ${\hat T_I} = \left( {{T_I} \circ {\cal F}} \right)\left( {{{\hat \psi }_{ab}}} \right)$. However, in many practical cases functions ${T_I}$ and/or ${\cal F}$ include division operations, and consequently their denominators may become zero under some values of ${\hat \psi _{ab}}$. Therefore, the existing adaptive observers designs \cite{b2, b3, b4, b5, b6, b7, b8, b9, b10, b11, b14} are not able to implement dynamic inversion-based change of coordinates $\hat x\left( t \right) = {\hat T_I}\left( t \right)\hat \xi \left( t \right)$ in {\it{bona fide}} manner and applicable only for the systems that are represented in the observer canonical form. The recent study \cite{b22} has paid great attention to solve this problem, but unfortunately the obtained results are not constructive for the systems that are not represented in the observer canonical form.}

So, the main objective of this technical note is to construct exponentially stable adaptive observers for systems that are not restricted to be represented in the observer canonical form. To achieve the above-mentioned goal, three key assumptions are introduced:
{
\begin{enumerate}
    \item[{\bf{A1.}}] A control signal ensures existence and boundedness of all system trajectories;
    \item [{\bf{A2.}}]All {physical} parameters of the system state space matrices are identifiable from parameters of numerator and denominator polynomials of the system input-output transfer function (identifiability criterion from \cite{b23, b24, b25} is met);
    \item [{\bf{A3.}}] The system is completely observable and desired pole placement equation for Luenberger correction feedback is known.
\end{enumerate}}

On the basis of the assumptions {\bf{A1}}-{\bf{A3}} and some sophisticated hypotheses an adaptive observer is proposed for linear time-invariant systems parameterized by unknown {physical} parameters with any form of representation of state space matrices. The procedure to design the proposed observer consists of four main steps:

\begin{enumerate}
    \item [{\bf{S1.}}] Following GPEBO \cite{b14} and DREM \cite{b15} design techniques, a measurable regression equation with scalar regressor with respect to parameters of denominator and numerator polynomials of system transfer function is obtained.
    \item [{\bf{S2.}}]Regression equation from {\bf{S1}} is transformed into division-free regression equation with respect to parameters of system matrices and Luenberger observer correction gain.
    \item [{\bf{S3.}}]Identification law from \cite{b26} is used to provide exponential convergence of estimates of system matrices unknown parameters and Luenberger observer correction gain under FE condition.
    \item [{\bf{S4.}}]Luenberger-type adaptive observer is applied to form the estimates of system states from estimates of unknown parameters of system matrices and Luenberger observer correction gain.
\end{enumerate}

When the assumptions {\bf{A1}}-{\bf{A3}} are met and the regressor is finitely exciting, the procedure {\bf{S1}}-{\bf{S4}} allows one to obtain adaptive observer that provides exponential convergence of the system parameter and state estimates to their true values. 

The remainder of the paper is organized as follows. Section II presents the rigorous problem statement. Section III provides some discussion on existing solutions for the problem under consideration. The proposed observer is elucidated in Section IV. Section V demonstrates the results of numerical experiments. The paper is wrapped up with the conclusion and goals of further research in Section VI.

\textbf{Notation and Definitions.} Further the following notation is used: $\left| . \right|$ is the absolute value, $\left\| . \right\|$ is the suitable norm of $(.)$, ${\lambda _{min }}\left( . \right)$ and ${\lambda _{max }}\left( . \right)$  are the matrix minimum and maximum eigenvalues respectively, ${\rm{vec}}\left( . \right)$ is the operation of a matrix vectorization, ${I_{n \times n}}=I_{n}$ is an identity $n \times n$ matrix, ${0_{n \times n}}$ is a zero $n \times n$ matrix, $0_{n}$ stands for a zero vector of length $n$, ${\rm{det}}\{.\}$ stands for a matrix determinant, ${\rm{adj}}\{.\}$ represents an adjoint matrix, $s$ is the variable of Laplace transform. For a mapping ${\cal F}{\rm{:\;}}{\mathbb{R}^n} \mapsto {\mathbb{R}^n}$  we denote its Jacobian by $\nabla_{x} {\cal F}\left( x \right) = \linebreak = {\textstyle{{\partial {\cal F}} \over {\partial x}}}\left( x \right)$. We also use the fact that for all (possibly singular) ${n \times n}$ matrices $M$ the following holds: ${\rm{adj}} \{M\} M = {\rm{det}} \{M\}I_{n \times n}$.

The definition of heterogeneous mapping, the regressor finite excitation condition, and the corollary of the Kalman-Yakubovich-Popov Lemma \cite{b29} are used in the paper.
\begin{definition}\label{definition1}
A mapping {$\mathcal{F}{\text{: \!}}{\mathbb{R}^{{n_x}}} \mapsto {\mathbb{R}^{{n_\mathcal{F}}\! \times\! {m_\mathcal{F}}}}$} is heterogeneous of degree ${\ell _\mathcal{F}}$ if there exists ${\Pi _\mathcal{F}}\left( {\omega} \right) \in {\mathbb{R}^{{n_\mathcal{F}} \times {n_\mathcal{F}}}}{\text{, }}$ ${\Xi _\mathcal{F}}\left( {\omega } \right)=$\linebreak$={\overline \Xi _\mathcal{F}}\left( {\omega } \right)\omega \in {\mathbb{R}^{{\Delta _\mathcal{F}} \times {n_x}}}$, and mapping ${\mathcal{T}_\mathcal{F}}{\text{: }} {\mathbb{R}^{{\Delta _\mathcal{F}}}} \mapsto {\mathbb{R}^{{n_\mathcal{F}} \times {m_\mathcal{F}}}}$ such that for all {$\omega  \in \mathbb{R}$}  and $x \in {\mathbb{R}^{{n_x}}}$ the following conditions are met:
\begin{equation}\label{eq1}
\begin{gathered}
  {\Pi _\mathcal{F}}\left( {\omega} \right)\mathcal{F}\left( x \right) = {\mathcal{T}_\mathcal{F}}\left( {{\Xi _\mathcal{F}}\left( {\omega} \right)x} \right){\text{, }} \\
  {\rm{det}}\left\{ {{\Pi _\mathcal{F}}}\left( {\omega} \right) \right\} \geqslant {\omega ^{{\ell _{_\mathcal{F}}}}}{\text{, }} \\
  {\Xi _\mathcal{F}}_{ij}\left( {\omega} \right) = {c_{ij}}{\omega ^{\ell_{ij}} }{\text{, }}
{{\overline \Xi }_{{\cal F}ij}}\left( \omega  \right) = {c_{ij}}{\omega ^{{\ell _{ij}} - 1}},\;{c_{ij}} \in \left\{ {0, \; 1} \right\}{\rm{,}}
\end{gathered}
\end{equation}
with ${\ell _\mathcal{F}} \geqslant 1,{\text{\;}}{\ell_{ij}}  \geqslant 1$.
\end{definition}

{Heterogeneous mapping is a generalization of a well-known homogeneous mapping. For example, the homogeneous mapping ${\cal F}\left( x \right) =\linebreak={x_1}{x_2}$ satisfies the condition \eqref{eq1} with 
\begin{gather*}
{\ell _{\cal F}} = 2,\;{\Pi _{\cal F}}\left( \omega  \right)= {\omega ^2}{\rm{,}}\\\;{\Xi _{\cal F}}\left( \omega  \right) = {\rm{diag}}\left\{ {\omega {\rm{,\;}}\omega } \right\}{\rm{,\;}}{\overline \Xi _{\cal F}}\left( \omega  \right) = {\rm{diag}}\left\{ {1,{\rm{\;}}1} \right\} 
\end{gather*}
and ${{\cal T}_{\cal F}}\left( {{\Xi _{\cal F}}\left( \omega  \right)x} \right){\rm{:}} = {\cal F}\left( {\omega x} \right).$}

{On the other hand, the mapping ${\cal F}\left( x \right) = {x_1}{x_2} + {x_1}$ is not homogeneous, but also satisfies definition 1 with
\begin{gather*}
{\ell _{\cal F}} = 2, {\Pi _{\cal F}}\left( \omega  \right) = {\omega ^2}{\rm{,\;}}{\Xi _{\cal F}}\left( \omega  \right) = {\begin{bmatrix}
\omega &0\\
0&\omega \\
{{\omega ^2}}&0
\end{bmatrix}}{\rm{,\;}}{\overline \Xi _{\cal F}}\left( \omega  \right) = {\begin{bmatrix}
1&0\\
0&1\\
\omega &0
\end{bmatrix}}
\end{gather*}
and some new function ${{\cal T}_{\cal F}}\left( {{\Xi _{\cal F}}\left( \omega  \right)x} \right) = \left( {\omega {x_1}} \right)\left( {\omega {x_2}} \right) + {\omega ^2}{x_1}$. }

{The definition is to determine the class of functions, for which a regression equation $y\left( t \right) = \omega \left( t \right)x$ with both measurable regressand $y(t)$ and a scalar regressor $\omega \left( t \right) \in \mathbb{R}$ can be transformed into a new equation with respect to ${\cal F}\left( x \right)$. Indeed, using the properties of \eqref{eq1}, we have ${\Pi _{\cal F}}\left( \omega  \right){\cal F}\left( x \right) = {{\cal T}_{\cal F}}\left( {{{\overline \Xi }_{\cal F}}\left( \omega  \right)y} \right)$. Such a transformation makes it possible to identify parameters of ${\cal F}\left( x \right)$ without application of substitution ${\cal F}\left( {\hat x} \right)$.}

\begin{definition}\label{definition2}
The regressor $\omega \left( t \right)$ is finitely exciting $\left( \omega \left( t \right) \in {\rm{FE}}\right)$ over the time range $\left[ {t_r^ + {\text{; }}{t_e}} \right]{\text{,}}$ if there exists $\alpha>0$ ,  $t_r^ +  \geqslant t_{0} \geqslant 0$, ${t_e} > t_r^ + $  such that the following inequality holds:
\begin{equation}\label{eq2}
\begin{gathered}
  \int\limits_{t_r^ + }^{{t_e}} {\omega \left( \tau  \right){\omega ^{\text{\rm T}}}\left( \tau  \right)d} \tau  \geqslant \alpha I_{n}{\text{,}}
\end{gathered}
\end{equation}
where $\alpha$ is the excitation level.
\end{definition}

\begin{corollary}
For any $D > 0$, controllable pair $\left( {A{\text{, }}B} \right)$ with \linebreak $B \in {\mathbb{R}^{n \times m}}$ and Hurwitz matrix $A \in {\mathbb{R}^{n \times n}}$ there exist matrices $P = {P^{\text{\rm T}}} > 0$, $Q \in {\mathbb{R}^{n \times m}},\;K \in {\mathbb{R}^{m \times m}}$ and a scalar $\mu  > 0$ such that:
\begin{equation}\label{eq3}
\begin{gathered}
  {A^{\text{\rm T}}}P + PA =  - Q{Q^{\text{\rm T}}} - \mu P{\text{, }}PB = QK{\text{, }} \\ 
  {K^{\text{\rm T}}}K = D + {D^{\text{\rm T}}}. \\ 
\end{gathered}
\end{equation}
\end{corollary}

\section{Problem Statement}

The following class of continuous linear systems parameterized
by unknown physical parameters $\theta$ is considered for all $t \geqslant {t_0}$:
{\begin{equation}\label{eq4}
\begin{array}{l}
\dot x\left( t \right) = A\left( \theta  \right)x\left( t \right) + B\left( \theta  \right)u\left( t \right) = {\Phi ^{\text{T}}}\left( {x{\text{, }}u} \right){\Theta _{AB}}\left( \theta  \right){\text{,}}\\
y\left( t \right) = {C^{\rm{T}}}x\left( t \right){\rm{,}}    
\end{array}
\end{equation}}
where
\begin{displaymath}
\begin{gathered}
  {\Phi ^{\text{T}}}\left( {x{\text{, }}u} \right) = {\begin{bmatrix}
  {{I_n} \otimes {x^{\text{T}}}\left( t \right)}&{{I_n} \otimes {u^{\text{T}}}\left( t \right)} 
\end{bmatrix}}{\mathcal{D}_\Phi } \in {\mathbb{R}^{n \times {n_\Theta }}}{\text{, }} \\ 
  {\Theta _{AB}}\left( \theta  \right) = {\mathcal{L}_\Phi } {\begin{bmatrix}
  {{\rm{vec}}^{\text{T}}\left( {{A^{\text{T}}}\left( \theta  \right)} \right)}&{B^{\text{T}}\left( \theta  \right)} 
\end{bmatrix}}^{\text{T}} \in {\mathbb{R}^{{n_\Theta }}}{\text{,}} 
\end{gathered}
\end{displaymath}
and $x\left( t \right) \in {\mathbb{R}^n}$ are physical states of the system with unknown initial conditions ${x_0}$, $u\left( t \right) \in \mathbb{R}$ is a control signal, $y\left( t \right) \in \mathbb{R}$ is an output, ${\Theta _{AB}} \in {\mathbb{R}^{{n_\Theta }}}{\text{, }}\theta  \in {D_\theta } \subset {\mathbb{R}^{{n_\theta }}}$ are unknown vectors such that ${n_\Theta } \geqslant {n_\theta }$, ${\mathcal{D}_\Phi } \in {\mathbb{R}^{\left( {{n^2} + n} \right) \times {n_\Theta }}}{\text{, }}$ ${\mathcal{L}_\Phi } \in {\mathbb{R}^{{n_\Theta } \times \left( {{n^2} + n} \right)}}$ are known duplication and elimination matrices, the vector ${C} \in {\mathbb{R}^n}$ and mapping ${\Theta _{AB}}{\text{:\;}}{\mathbb{R}^{{n_\theta }}} \!\!\mapsto\!\! {\mathbb{R}^{{n_\Theta }}}$ are known\footnote{If it does not cause confusion, further we will occasionally omit dependence from $\theta$  {and/or $t$} for the sake of brevity.}. Only $u\left( t \right)$ and $y\left( t \right)$ are available for measurement, and the following classical assumptions are supposed to be met for system matrices and control signal.

\begin{assumption}
The control signal $u\left( t \right)$ ensures existence and boundedness of all trajectories of systems \eqref{eq4} for all $t \geqslant {t_0}$.
\end{assumption}
\begin{assumption}{
In the whole domain ${D_\theta}$ the parameters $\theta$ are identifiable via input/output signals, i.e. for all $\theta  \in {D_\theta }$  the Jacobian of Markov parameters matrix have a full rank:
\begin{equation}\label{eq5}
G\left( \theta  \right) = {\begin{bmatrix}
{{C^{\rm{T}}}B\left( \theta  \right)}\\
 \vdots \\
{{C^{\rm{T}}}{A^{2n - 1}}\left( \theta  \right)B\left( \theta  \right)}
\end{bmatrix}}{\rm{,\;}}rank\left\{ {{\nabla _\theta }G\left( \theta  \right)} \right\} = {n_\theta }{\rm{.}}
\end{equation}}
\end{assumption}
\begin{assumption}{
In the whole domain ${D_\theta}$ the pair $\left( {{C^{\rm{T}}}{\rm{,\;}}A\left( \theta  \right)} \right)$ is observable.}
\end{assumption}
The goal is to reconstruct unmeasured states $x(t)$ with the help of following adaptive observer:
\begin{equation}\label{eq6}
\begin{array}{l}
\dot {\hat {x}}\left( t \right) = {\Phi ^{\rm{T}}}\left( {\hat x{\rm{,\;}}u} \right){{\hat \Theta }_{AB}}\left( t \right) - \hat L\left( t \right)\left( {\hat y\left( t \right) - y\left( t \right)} \right){\rm{,}}\\
\hat y\left( t \right) = {C^{\rm{T}}}\hat x\left( t \right){\rm{,}}
\end{array}
\end{equation}
where ${\hat \Theta _{AB}}\left( t \right)$ is a system parameters estimate, $\tilde y\left( t \right) = \hat y\left( t \right) - y\left( t \right)$ denotes the output observation error, $L{\rm{:\;}}{D_\theta } \mapsto {D_L}$ is a known mapping to calculate a Luenberger correction gain via {pole placement $\det \left\{ {s{I_n} - A\left( \theta  \right) + L\left( \theta  \right){C^{\rm{T}}}} \right\} = \det \left\{ {s{I_n} - {A_{ref}}} \right\}$ with some known Hurwitz matrix ${A_{ref}} \in {\mathbb{R}^{n \times n}}$}, $\hat L\left( t \right) \in {\mathbb{R}^n}$ stands for the estimate of the Luenberger correction gain $L\left( \theta  \right)$.

The observer \eqref{eq6}  needs to be augmented with the identification laws, which under finite excitation condition ensures that the following objective is achieved:
\begin{equation}\label{eq7}
\begin{array}{c}
\mathop {{\text{lim}}}\limits_{t \to \infty } \left\| {\tilde x\left( t \right)} \right\| = 0{\text{ }}\left( {\exp } \right){\text{, }}\\\mathop {{\text{lim}}}\limits_{t \to \infty } \left\| {{{\tilde \Theta }_{AB}}\left( t \right)} \right\| = 0{\text{ }}\left( {\exp } \right){\text{,\;}}\mathop {{\rm{lim}}}\limits_{t \to \infty } \left\| {\tilde L\left( t \right)} \right\| = 0{\rm{\;}}\left( {\exp } \right){\rm{,}}
\end{array}
\end{equation}
where $\tilde x\left( t \right) = \hat x\left( t \right) - x\left( t \right)$ is the state observation error, \linebreak ${\tilde \Theta _{AB}}\left( t \right) = {\hat \Theta _{AB}}\left( t \right) - {\Theta _{AB}}\left( \theta  \right)$ and  $\tilde L\left( t \right) = \hat L\left( t \right) - L\left( \theta  \right)$ are the parametric errors, { $\exp$ is the abbreviation for the exponential rate of convergence.}

\section{Preliminaries}
Before introduction of the proposed adaptive laws, it is worth discussing some existing approaches to solve the problem under consideration. {Many methods, e.g. \cite{b2, b3, b4, b5, b6, b7, b8, b9, b10, b11, b14}, first of all, apply a transformation of the original system to the observer canonical form. Indeed, if Assumption 3 is met, in all domain ${D_\theta }$  the system \eqref{eq4} can be transformed into the observer canonical form:
\begin{equation}\label{eq8}
\begin{array}{l}
\dot \xi \left( t \right) = {A_0}\xi \left( t \right) + {\psi _a}\left( \theta  \right)y\left( t \right) + {\psi _b}\left( \theta  \right)u\left( t \right){\rm{,}}\\
y\left( t \right) = C_0^{\rm{T}}\xi \left( t \right){\rm{,\;}}\xi \left( {{t_0}} \right) = {\xi _0}\left( \theta  \right),
\end{array}
\end{equation}
where
\begin{gather*}
\begin{array}{c}
{\psi _a}\left( \theta  \right) = T\left( \theta  \right)A\left( \theta  \right){T^{ - 1}}\left( \theta  \right){C_0}{\rm{,\;}}{\psi _b}\left( \theta  \right) = T\left( \theta  \right)B\left( \theta  \right){\rm{, }}\\
{A_0} = {\begin{bmatrix}
{{0_n}}&{\begin{matrix}
{{I_{n - 1}}}\\
{{0_{1 \times \left( {n - 1} \right)}}}
\end{matrix}}
\end{bmatrix}}{\rm{,\;}}\begin{array}{*{20}{c}}
{C_0^{\rm{T}} = {C^{\rm{T}}}{T^{ - 1}}\left( \theta  \right)}\\
{\xi \left( {{t_0}} \right) = {T^{ - 1}}\left( \theta  \right){x_0}}
\end{array},\\
{T_I}\left( \theta  \right){\rm{:}} = {T^{ - 1}}\left( \theta  \right) = {\begin{bmatrix}
{{A^{n - 1}}\left( \theta  \right){{\cal O}_n}}&{{A^{n - 2}}\left( \theta  \right){{\cal O}_n}}& \cdots &{{{\cal O}_n}}
\end{bmatrix}}{\rm{,}}
\end{array}  
\end{gather*}
and $\xi \left( t \right) \in {\mathbb{R}^n}$ are virtual states of the system, ${\psi _a}{\rm{,\;}}{\psi _b}{\rm{:\;}}{D_\theta } \mapsto {D_\psi }$ are known differentiable functions, ${{\cal O}_n}$ is the $n^{th}$ column of the matrix that is an inverse one to ${{\cal O}^{ - 1}} =\linebreak= {{\begin{bmatrix}
{{C^{\rm{T}}}}&{{C^{\rm{T}}}A\left( \theta  \right)}& \cdots &{{C^{\rm{T}}}{A^{n - 1}}\left( \theta  \right)}
\end{bmatrix}}^{\rm{T}}}.$}

It is well known that in general case using measurable $u\left( t \right){\rm{,\;}}y\left( t \right)$ and model \eqref{eq8}, only the parameters ${\psi _a}{\rm{,\;}}{\psi _b}$ of numerator/denominator polynomials of the transfer function ${C^{\rm{T}}}{\left( {sI - A\left( \theta  \right)} \right)^{ - 1}}B\left( \theta  \right)$ are identifiable \cite{b23, b24, b25, b29}.  However, in the considered overparameterized case the parameters ${\Theta _{AB}}$ of the system depend nonlinearly on the physical parameters $\theta$. In their turn, the parameters ${\psi _a}{\rm{,\;}}{\psi _b}$ of numerator and denominator polynomials of the transfer function ${C^{\rm{T}}}{\left( {sI - A\left( \theta  \right)} \right)^{ - 1}}B\left( \theta  \right)$ also depend on $\theta$ in nonlinear manner. Therefore, when ${\psi _a}{\rm{,\;}}{\psi _b}$ are differentiable and in the whole domain ${D_\psi }$ for some their elements (handpicked by matrix ${{\cal L}_{ab}} \in {\mathbb{R}^{{n_\theta } \times 2n}}$) the following condition is met:
\begin{equation}\label{eq9}
\begin{array}{c}
{\det ^2}\left\{ {{\nabla _\theta }{\psi _{ab}}\left( \theta  \right)} \right\} > 0,{\rm{ }}\\
{\psi _{ab}}\left( \theta  \right) = {{\cal L}_{ab}}{\begin{bmatrix}
{{\psi _a}\left( \theta  \right)}\\
{{\psi _b}\left( \theta  \right)}
\end{bmatrix}}\in {\mathbb{R}^{{n_\theta }}}{\rm{,}}
\end{array}
\end{equation}
then, according to the inverse function theorem, there exists an inverse mapping ${\cal F}{\rm{:\;}}{D_\psi } \mapsto {D_\theta }$ such that $\theta  = {\cal F}\left( {{\psi _{ab}}} \right)$. {Moreover, it should be mentioned that, according to the structural identifiability criterion \cite{b23, b24, b25}, if \eqref{eq9} is not verified, then effective reconstruction of the parameters  $\theta$  via input/output data will be impossible, and therefore the stated goal \eqref{eq7} is unachievable. }
\begin{remark}
{Equivalence of \eqref{eq5} and \eqref{eq9} has been thoroughly discussed in the related studies \cite{b23, b24, b25}, which means that, if \eqref{eq5} is met, then there exists a matrix ${{\cal L}_{ab}} \in {\mathbb{R}^{{n_\theta } \times 2n}}$ such that \eqref{eq9} is also met and the converse is also true. }
\end{remark}
{Consequently, for the class of structurally identifiable systems, in the domain ${D_\psi }$ it becomes possible to: {\it{i}}) find the system parameters $\theta$ from ${\psi _{ab}}$, {\it{ii}}) recalculate $\theta$ into ${\Theta _{AB}}\left( \theta  \right){\rm{,\;}}{T_I}\left( \theta  \right){\rm{,\;}}L\left( \theta  \right)$ etc., \linebreak {\it{iii}}) design an adaptive observers to form estimates of the original states $\hat x\left( t \right)$. A well-known approach to implement these procedures under condition that the system parameters ${\psi _{ab}}\left( \theta  \right)$ are unknown is to use the following {\it{certainty equivalence}} substitutions:
\renewcommand{\theequation}{\arabic{equation}a}
\begin{equation}\label{eq10}
\begin{array}{l}
{{\hat \Theta }_{AB}}\left( t \right) = \left( {{\Theta _{AB}} \circ {\cal F}} \right)\left( {{{\hat \psi }_{ab}}\left( t \right)} \right){\rm{,}}\\
{{\hat T}_I}\left( t \right) = \left( {{T_I} \circ {\cal F}} \right)\left( {{{\hat \psi }_{ab}}\left( t \right)} \right){\rm{, }}\\
\hat x\left( t \right) = {{\hat T}_I}\left( t \right)\hat \xi \left( t \right){\rm{,}}\\
\end{array}
\end{equation}
or
\setcounter{equation}{9}
\renewcommand{\theequation}{\arabic{equation}b}
\begin{equation}
\begin{array}{l}
{{\hat \Theta }_{AB}}\left( t \right) = \left( {{\Theta _{AB}} \circ {\cal F}} \right)\left( {{{\hat \psi }_{ab}}\left( t \right)} \right){\rm{,}}\\
\hat L\left( t \right) = \left( {L \circ {\cal F}} \right)\left( {{{\hat \psi }_{ab}}\left( t \right)} \right){\rm{,}}\\
\dot {\hat {x}}\left( t \right) = {\Phi ^{\rm{T}}}\left( {\hat x{\rm{, }}u} \right){{\hat \Theta }_{AB}}\left( t \right) - \hat L\left( t \right)\left( {\hat y\left( t \right) - y\left( t \right)} \right){\rm{,}} 
\end{array}
\end{equation}
where the estimates ${\hat \psi _{ab}}\left( t \right)$ and $\hat \xi \left( t \right)$ could be obtained under appropriate excitation conditions using any of the existing adaptive observers designs \cite{b2, b3, b4, b5, b6, b7, b8, b9, b10, b11, b14} for the systems in observer canonical form \eqref{eq8}.}
\renewcommand{\theequation}{\arabic{equation}}
\setcounter{equation}{10}

{However, dynamic substitution (10) is singular if the signals ${\hat \psi _a}\left( t \right){\rm{,\;}}{\hat \psi _b}\left( t \right)$ leave their ”safe” domain ${D_\psi }$ of the system identifiability. Therefore, the existing adaptive observers designs \cite{b2, b3, b4, b5, b6, b7, b8, b9, b10, b11, b14} are not able to implement dynamic inversion-based change of coordinates in {\it{bona fide}} manner and consequently applicable only for the systems that are represented in the observer canonical form ($A\left( \theta  \right){\rm{,\;}}B\left( \theta  \right)$ in \eqref{eq4} must coincide with ${A_0} + {\psi _a}\left( \theta  \right){\rm{,\;}}{\psi _b}\left( \theta  \right)$ in \eqref{eq8}, respectively). Procedures of parameters projection do not provide a good solution for this drawback because domain ${D_\psi }$ can be nonconvex and have very complex geometry.}

In this technical note, we make some effort to solve above-mentioned problem and propose a novel adaptive observer, which is applicable in terms of the goal \eqref{eq7} for the structurally identifiable completely observable systems that are not necessarily represented in the observer canonical form \eqref{eq8}. To facilitate the design procedure under condition that criterion \eqref{eq5}, \eqref{eq9} is satisfied and ${{\cal L}_{ab}} \in {\mathbb{R}^{{n_\theta } \times 2n}}$ is known, we additionally adopt the following hypotheses.

\begin{Hypothesis} 
The mapping ${\Theta _{AB}}{\rm{:\;}}{D_\theta } \mapsto {D_\Theta }$ is heterogeneous in the sense of \eqref{eq1} such that:
\begin{equation}\label{eq11}
{\Pi _\Theta }\left({{{\cal M}_\theta}} \right){\Theta _{AB}}\left( \theta  \right) = {\mathcal{T}_\Theta }\left( {{\Xi _\Theta }\left({{{\cal M}_\theta}}\right)\theta } \right){\text{,}}
\end{equation}
where $\det \left\{ {{\Pi _\Theta }\left( {{{\cal M}_\theta }} \right)} \right\} \ge {\cal M}_\theta ^{{\ell _\Theta }}{\text{, }}{\ell _\Theta } \geqslant 1$, ${\Xi _\Theta }\left({{{\cal M}_\theta }} \right) \in {\mathbb{R}^{{\Delta _\Theta } \times {n_\theta }}}$, ${{\cal T}_\Theta }{\rm{:\;}}{\mathbb{R}^{{\Delta _\Theta }}} \mapsto {\mathbb{R}^{{n_\Theta }}}{\rm{,}}$ and all mappings are known.
\end{Hypothesis}

\begin{Hypothesis} 
There exist heterogeneous in the sense of \eqref{eq1} mappings $\mathcal{G}{\rm{:\;}}{D_\psi }\mapsto {\mathbb{R}^{{n_\theta } \times {n_\theta }}}$, $\mathcal{S}{\rm{:\;}}{D_\psi }\mapsto {\mathbb{R}^{{n_\theta }}}$ such that:
\begin{equation}\label{eq12}
\begin{gathered}
  \mathcal{S}\left( {{\psi _{ab}}} \right) = \mathcal{G}\left( {{\psi _{ab}}} \right)\mathcal{F}\left( {{\psi _{ab}}} \right) = \mathcal{G}\left( {{\psi _{ab}}} \right)\theta {\text{,}} \\ 
  {\Pi _\theta }\left( {\Delta} \right)\mathcal{G}\left( {{\psi _{ab}}} \right)\!\! =\!\! {\mathcal{T}_\mathcal{G}}\left( {{\Xi _\mathcal{G}}\left( {\Delta} \right){\psi _{ab}}} \right){\text{,}} \\ 
  {\Pi _\theta }\left( {\Delta} \right)\mathcal{S}\left( {{\psi _{ab}}} \right) = {\mathcal{T}_\mathcal{S}}\left( {{\Xi _\mathcal{S}}\left( {\Delta} \right){\psi _{ab}}} \right){\text{,}} \\ 
\end{gathered} 
\end{equation}
where $\det \left\{ {{\Pi _\theta }\left( {\Delta} \right)} \right\} \geqslant {\Delta ^{{\ell _\theta }}}\left( t \right){\text{, }}{\rm{rank}}\left\{ {\mathcal{G}\left( {{\psi _{ab}}} \right)} \right\} = {n_\theta }{\text{, }}{\ell _\theta } \geqslant 1$, ${\Xi _\mathcal{G}}\left( {\Delta} \right) \in {\mathbb{R}^{{\Delta _\mathcal{G}} \times {n_\theta }}}$, ${\Xi _\mathcal{S}}\left( {\Delta} \right) \in {\mathbb{R}^{{\Delta _\mathcal{S}} \times {n_\theta }}}$, ${{\cal T}_{\cal G}}{\rm{:\;}}{\mathbb{R}^{{\Delta _{\cal G}}}} \mapsto {\mathbb{R}^{{n_\theta } \times {n_\theta }}}{\rm{,\;}}\linebreak{{\cal T}_{\cal S}}{\rm{:\;}}{\mathbb{R}^{{\Delta _{\cal S}}}} \mapsto {\mathbb{R}^{{n_\theta }}}$ and all mappings are known.
\end{Hypothesis}

\begin{Hypothesis}
{There exist heterogeneous in the sense of \eqref{eq1} mappings ${\cal Q}{\rm{:\;}}{D_\psi } \mapsto {\mathbb{R}^n}{\rm{,\;}}{\cal P}{\rm{:\;}}{D_\psi } \mapsto {\mathbb{R}^{n \times n}}$  such that:
\begin{equation}\label{eq13}
    \begin{array}{c}
{\cal Q}\left( \theta  \right) = {\cal P}\left( \theta  \right)L\left( \theta  \right){\rm{,}}\\
{\Pi _L}\left( {{{\cal M}_\theta }} \right){\cal P}\left( \theta  \right) = {{\cal T}_{\cal P}}\left( {{\Xi _{\cal P}}\left( {{{\cal M}_\theta }} \right)\theta } \right){\rm{,}}\\
{\Pi _L}\left( {{{\cal M}_\theta }} \right){\cal Q}\left( \theta  \right) = {{\cal T}_{\cal Q}}\left( {{\Xi _{\cal Q}}\left( {{{\cal M}_\theta }} \right)\theta } \right){\rm{,}}
\end{array}
\end{equation}
where $\det \left\{ {{\Pi _L}\left( {{{\cal M}_\theta }} \right)} \right\} \ge {\omega ^{{\ell _L}}}\left( t \right){\rm{,\;}}{\rm{rank}}\left\{ {{\cal P}\left( \theta  \right)} \right\} = n{\rm{,\;}}{\ell _L} \ge 1$, ${\Xi _{\cal P}}\left( {{{\cal M}_\theta }} \right) \in {\mathbb{R}^{{\Delta _{\cal P}} \times {n_\theta }}}$, ${\Xi _{\cal Q}}\left( {{{\cal M}_\theta }} \right) \in {\mathbb{R}^{{\Delta _{\cal Q}} \times {n_\vartheta }}}{\rm{,}}$ ${{\cal T}_{\cal P}}{\rm{:\;}}{\mathbb{R}^{{\Delta _{\cal P}}}} \mapsto {\mathbb{R}^n}{\rm{,\;}}\linebreak{{\cal T}_{\cal Q}}{\rm{:\;}}{\mathbb{R}^{{\Delta _{\cal Q}}}} \mapsto {\mathbb{R}^n}$  and all mappings are known.}
\end{Hypothesis}

{Some intuition about the role of the above-mentioned hypotheses is discussed here on the basis of a simple example. Let $\theta  =\linebreak= {\cal F}\left( {{\psi _{ab}}} \right) = col\left\{ {{\psi _{2ab}}\psi _{1ab}^{ - 1}{\rm{,\;}}\sqrt {{\psi _{1ab}}}  + \psi _{1ab}^2} \right\}$ with ${\psi _{1ab}} > 0$, the signals ${\cal Y}\left( t \right) \in {\mathbb{R}^{{n_\theta }}}$ and $\Delta \left( t \right) \in \mathbb{R}$ be measurable and satisfy the regression equation ${\cal Y}\left( t \right) = \Delta \left( t \right){\psi _{ab}}\left( \theta  \right)$. Using Hypothesis 2, ${\cal F}\left( {{\psi _{ab}}} \right)$ is factorized as follows:}
{\begin{gather*}
    {\cal S}\left( {{\psi _{ab}}} \right) =  {\begin{bmatrix}
{{\psi _{2ab}}}\\
{\sqrt {{\psi _{1ab}}}  + \psi _{1ab}^2}
\end{bmatrix}} {\rm{,\;}}{\cal G}\left( {{\psi _{ab}}} \right) = {\begin{bmatrix}
{{\psi _{1ab}}}&0\\
0&1
\end{bmatrix}}{\rm{,}}
\end{gather*}
then, choosing ${\Pi _\theta }\left( \Delta  \right) = diag\left\{ {\Delta {\rm{,\;}}{\Delta ^2}} \right\}$, it is written:
\begin{gather*}
\underbrace {{\begin{bmatrix}
\Delta &0\\
0&{{\Delta ^2}}
\end{bmatrix}} {\begin{bmatrix}
{{\psi _{2ab}}}\\
{\sqrt {{\psi _{1ab}}}  + \psi _{1ab}^2}
\end{bmatrix}}}_{{{\cal T}_{\cal S}}\left( {{\Xi _{\cal S}}\left( \Delta  \right){\psi _{ab}}} \right)} = \underbrace { {\begin{bmatrix}
\Delta &0\\
0&{{\Delta ^2}}
\end{bmatrix}}{\begin{bmatrix}
{{\psi _{1ab}}}&0\\
0&1
\end{bmatrix}}}_{{{\cal T}_{\cal G}}\left( {{\Xi _{\cal G}}\left( \Delta  \right){\psi _{ab}}} \right)}\theta, 
\end{gather*}
where ${\Xi _{\cal S}}\left( \Delta  \right) = {\begin{bmatrix}
\Delta &0\\
{{\Delta ^4}}&0\\
0&\Delta 
\end{bmatrix}}{\rm{,\;}}{\Xi _{\cal G}}\left( \Delta  \right) = {\begin{bmatrix}
\Delta &0
\end{bmatrix}}$ and ${\Delta ^2}\left( t \right)$ in the second equation is tractable as a free term.}

{As ${\cal Y}\left( t \right) = \Delta \left( t \right){\psi _{ab}}\left( \theta  \right)$ and $\Delta \left( t \right)$ are measurable, then, owing to ${\psi _{1ab}} > 0,{\rm{\;}}{\Delta^2\left( t \right)}\sqrt {{\psi _{1ab}}}  = \sqrt {{\Delta^4\left( t \right)}{\psi _{1ab}}} $ , the following redefinition holds:
\begin{gather*}
\underbrace {{\begin{bmatrix}
{{{\cal Y}_2}\left( t \right)}\\
{\sqrt {{\Delta ^3}\left( t \right){{\cal Y}_1}\left( t \right)}  + {\cal Y}_1^2\left( t \right)}
\end{bmatrix}}}_{{{\cal T}_{\cal S}}\left( {{{\overline \Xi }_{\cal S}}\left( \Delta  \right){\cal Y}} \right)} = \underbrace {{\begin{bmatrix}
{{{\cal Y}_1}\left( t \right)}&0\\
0&{{\Delta ^2}\left( t \right)}
\end{bmatrix}}}_{{{\cal T}_{\cal G}}\left( {{{\overline \Xi }_{\cal G}}\left( \Delta  \right){\cal Y}} \right)}\theta {\rm{,}}
\end{gather*}
where ${\overline \Xi _{\cal S}}\left( \Delta  \right) = {\begin{bmatrix}
1&0\\
{{\Delta ^3}}&0\\
0&1
\end{bmatrix}}{\rm{,\;}}{\overline \Xi _{\cal G}}\left( \Delta  \right) = {\begin{bmatrix}
1&0
\end{bmatrix}}.$}

{The fact that conditions ${\rm{det}} \left\{ {{\Pi _\theta }\left( \Delta  \right)} \right\} \ge {\Delta ^{{\ell _\theta }}}{\rm{,\;}}{\ell _\theta } \ge 1$, ${\rm{rank}}\left\{ {{\cal G}\left( {{\psi _{ab}}} \right)} \right\} = {n_\theta }$  hold is ensured by the definition of ${\cal F}\left( {{\psi _{ab}}} \right)$ and chosen matrix ${\Pi _\theta }\left( \Delta  \right)$.}

{The signals ${{\cal T}_{\cal S}}\left( {{{\overline \Xi }_{\cal S}}\left( \Delta  \right){\cal Y}} \right)$ and ${{\cal T}_{\cal G}}\left( {{{\overline \Xi }_{\cal G}}\left( \Delta  \right){\cal Y}} \right)$ are known (can be computed), and hence a simple gradient law to identify the parameters $\theta$ can be derived from the obtained equation \cite{b29}. In contrast to the {\it{certainty equivalence}} based recalculation approach (10), such estimates $\hat \theta \left( t \right)$ are always singularity free as: {\it{i}}) mappings ${{\cal T}_{\cal S}}$ and ${{\cal T}_{\cal G}}$ do not include division operations because of the decomposition \eqref{eq12}, and {\it{ii}}) according to the hypotheses, the domains of these mappings are the whole spaces ${\mathbb{R}^{{\Delta _{\cal G}}}}$ and ${\mathbb{R}^{{\Delta _{\cal S}}}}$, respectively. At the same time, for example, if the {\it{certainty-equivalence}} based approach (10) is applied, then the estimates ${\hat \psi _{ab}}\left( t \right)$  are explicitly substituted into the mapping ${\cal F}$ to obtain:
\begin{gather*}
    \hat \theta \left( t \right) = {\cal F}\left( {{{\hat \psi }_{ab}}} \right) = col\left\{ {{\textstyle{{{{\hat \psi }_{2ab}}} \over {{{\hat \psi }_{1ab}}}}}{\rm{, }}\sqrt {{{\hat \psi }_{1ab}}}  + \hat \psi _{1ab}^2} \right\}{\rm{,}}
\end{gather*}
and a singularity occurs when ${\hat \psi _{1ab}}\left( t \right) \le 0$.}

Therefore, Hypotheses 1 and 3 describe the conditions, under which we can obtain the linear regression equations (LRE) with respect to ${\Theta _{AB}}\left( \theta  \right)$ and $L\left( \theta  \right)$ from the regression equation ${{\cal Y}_\theta }\left( t \right) = \linebreak = {{\cal M}_\theta }\left( t \right)\theta$, where ${{\cal M}_\theta }\left( t \right) \in \mathbb{R}$. Hypothesis 2 sets the conditions to obtain LRE with respect to $\theta$ from ${\cal Y}\left( t \right) = \Delta \left( t \right){\psi _{ab}}\left( \theta  \right)$, where $\Delta \left( t \right) \in \mathbb{R}$. As compared to 2 and 3, Hypothesis 1 requires that the parameters ${\Theta _{AB}}\left( \theta  \right)$ are evaluated without division operations. However, if necessary, this assumption can be reformulated in terms of \eqref{eq12} and \eqref{eq13}.

\begin{remark} 
Hypotheses 1-3 are not constructive, since the ways of how to obtain the above-mentioned mappings are not considered, but only assumed to exist and be known. However, it is easy to see that condition \eqref{eq12} is satisfied for all algebraic function. In many practical applications Hypotheses 1-3 are reasonable and not restrictive.
\end{remark}

\section{Main Result}

In case Hypotheses \eqref{eq11}-\eqref{eq13} hold, the main result of this study is a set of techniques to: ({\it i}) compute the function ${\cal Y}\left( t \right)$, which is then used to ({\it ii}) obtain regression equations with respect to ${\Theta _{AB}}\left( \theta  \right)$ and $L\left( \theta  \right)$ that are the basis to ({\it iii}) derive identification laws, in their turn, ensuring that the goal \eqref{eq7} is met. Towards this end, first of all, the parametrization from PEBO procedure is applied to obtain a set of regression equations with respect to unknown parameters ${\psi _{ab}}\left( \theta  \right)$.

\begin{lemma} 
The unknown parameters ${\psi _{ab}}\left( \theta  \right)$ satisfy the following linear regression model:
\begin{equation}\label{eq14}
\begin{gathered}
  \mathcal{Y}\left( t \right) = {\mathcal{L}_{ab}}{\mathcal{L}_0}\Delta \left( t \right)\eta (\theta)  = \Delta \left( t \right){\psi _{ab}}\left( \theta  \right){\text{,}} \\ 
  {\mathcal{Y}\left( t \right) = k \cdot {\mathcal{L}_{ab}}{\mathcal{L}_0}{\rm{adj}}\left\{ {\overline \varphi \left( t \right)} \right\}\overline q\left( t \right){\text{,}}}\;{\Delta \left( t \right) = k \cdot {\rm{det}}\left\{ {\overline \varphi \left( t \right)} \right\}{\text{,}}} 
\end{gathered}
\end{equation}
where
\begin{equation}\label{eq15}
\begin{gathered}
  {\dot {\overline q}}\left( t \right) = {e^{ - \sigma \left( {t - {t_0}} \right)}}\varphi \left( t \right)q\left( t \right) \in {\mathbb{R}^{3n}}{\text{, }}\overline q\left( {{t_0}} \right) = {0_{3n}}, \\ 
  {\dot {\overline \varphi}} \left( t \right) = {e^{ - \sigma \left( {t - {t_0}} \right)}}\varphi \left( t \right){\varphi ^{\text{\rm T}}}\left( t \right) \in {\mathbb{R}^{3n \times 3n}}{\text{, }}\overline \varphi \left( {{t_0}} \right) = {0_{3n \times 3n}}, \\ 
\end{gathered}
\end{equation}

\begin{equation}\label{eq16}
\begin{gathered}
q\left( t \right)\! =\! y - C_0^{\text{\rm T}}z{\text{, }}\varphi \left( t \right) = {\begin{bmatrix}
  {{\Omega ^{\text{\rm T}}}{C_0}} \\ 
  {{P^{\text{\rm T}}}{C_0}} \\ 
  { {e^{{A_K}\left( {t - {t_0}} \right)}}{C_0}} 
\end{bmatrix}} {\text{, }}\eta  = {\begin{bmatrix}
  {{\psi _a}\left( \theta  \right)} \\ 
  {{\psi _b}\left( \theta  \right)} \\ 
  {\tilde \xi _0 \left( {{\theta}} \right)} 
\end{bmatrix}} {\text{,}} \\ 
  \dot z\left( t \right) = {A_K}z\left( t \right) + Ky\left( t \right){\text{, }}z\left( {t^+_0 } \right) = {0_n}{\text{,}} \\ 
  \dot \Omega \left( t \right) = {A_K}\Omega \left( t \right) + {I_n}y\left( t \right){\text{, }}\Omega \left( {t^+_0 } \right) = {0_{n \times n}}{\text{,}} \\ 
  \dot P\left( t \right) = {A_K}P\left( t \right) + {I_n}u\left( t \right){\text{, }}P\left( {t^+_0 } \right) = {0_{n \times n}} \\ 
\end{gathered}
\end{equation}
and $\forall t \geqslant {t_e}{\text{ }}\Delta \left( t \right) \geqslant {\Delta _{{\rm{min}}}} > 0$ when $\varphi \left( t \right) \in {\rm{FE}}$, $\sigma  > 0$ is a damping ratio, $k \geqslant k_{min} > 0$ is an amplitude modulator, ${A_K} = {A_0} - KC_0^{\text{\rm T}}$ is a stable matrix, ${\mathcal{L}_0} = {\begin{bmatrix}
  {{I_{2n \times 2n}}}&{{0_{2n \times n}}} 
\end{bmatrix}} $ is an eliminator to convert $\eta(\theta)$ into ${\begin{bmatrix}
{{\psi _a}\left( \theta  \right)}\\
{{\psi _b}\left( \theta  \right)}
\end{bmatrix}} $.

Proof of Lemma 1 is postponed to Appendix.
\end{lemma}

Having the regression equation \eqref{eq14} at hand and using \eqref{eq11}, \eqref{eq12} and \eqref{eq13}, we are in position to obtain the regression equations with respect to $\theta {\text{, }}{\Theta _{AB}(\theta)}$ and $L(\theta)$.

\begin{lemma}
The unknown parameters $\theta {\text{, }}{\Theta _{AB}}\left( \theta  \right)$ and $L\left( \theta  \right)$ satisfy the following measurable linear regression models:
\begin{equation}\label{eq17}
{\mathcal{Y}_\theta }\left( t \right) = {\mathcal{M}_\theta }\left( t \right)\theta {\text{,}}
\end{equation}
\begin{equation}\label{eq18}
{\mathcal{Y}_{AB}}\left( t \right) = {\mathcal{M}_{AB}}\left( t \right){\Theta _{AB}}\left( \theta  \right){\text{,}}
\end{equation}
\begin{equation}\label{eq19}
{\mathcal{Y}_L}\left( t \right) = {\mathcal{M}_L}\left( t \right)L\left( \theta  \right){\text{,}}
\end{equation}	
where
\begin{equation}\label{eq20}
\begin{gathered}
  {\mathcal{Y}_{AB}}\left( t \right) = {\rm{adj}}\left\{ {{\Pi _\Theta }\left( {{\mathcal{M}_\theta }} \right)} \right\}{\mathcal{T}_\Theta }\left( {{{\overline \Xi }_\Theta }\left( {{\mathcal{M}_\theta }} \right){\mathcal{Y}_\theta }} \right){\text{,}} \\ 
  {\mathcal{Y}_L}\!\! \left( t \right) = {\rm{adj}} \! \left\{\! {{\mathcal{T}_\mathcal{P}}\left( {{{\overline \Xi }_\mathcal{P}}\left( {{\mathcal{M}_\theta }} \right){\mathcal{Y}_\theta }} \right)} \! \right\}\!{\mathcal{T}_\mathcal{Q}}\!\left( {{{\overline \Xi }_\mathcal{Q}}\left( {{\mathcal{M}_\theta }} \right){\mathcal{Y}_\theta } } \right){\text{,}} \\ 
  {\mathcal{Y}_\theta }\left( t \right) = {\rm{adj}}\left\{ \!{{\mathcal{T}_\mathcal{G}}\left( {{{\overline \Xi }_\mathcal{G}}\left( {\Delta } \right)\mathcal{Y}} \right)} \! \right\}{\! \mathcal{T}_\mathcal{S}}\!\left( {{{\overline \Xi }_\mathcal{S}}\left( {\Delta } \right)\mathcal{Y}} \right){\text{,}} \\  
  {\mathcal{M}_{AB}}\left( t \right) = {\rm{det}} \left\{ {{\Pi _\Theta }\left( {{\mathcal{M}_\theta }} \right)} \right\}{\text{, }}\\{\mathcal{M}_L}\left( t \right) = {\rm{det}} \left\{ {{\mathcal{T}_\mathcal{P}}\left( {{{\overline \Xi }_\mathcal{P}}\left( {{\mathcal{M}_\theta }} \right){\mathcal{Y}_\theta}} \right)} \right\}{\text{,}} \\ 
  {\mathcal{M}_\theta }\left( t \right) = {\rm{det}} \left\{ {{\mathcal{T}_\mathcal{G}}\left( {{{\overline \Xi }_\mathcal{G}}\left( {\Delta} \right)\mathcal{Y}} \right)} \right\}, \\
\end{gathered}
\end{equation}
and $\forall t \geqslant {t_e}{\text{ }}\left|{\mathcal{M}_{AB}}\left( t \right)\right| \geqslant \underline {{\mathcal{M}_{AB}}}  > 0,{\text{ }}\left|{\mathcal{M}_L}\left( t \right)\right| \geqslant \underline {{\mathcal{M}_L}}  > 0,{\text{ }}$ when $\varphi \left( t \right) \in {\rm{FE}}$.

Proof of Lemma 2 is presented in Appendix.
\end{lemma}

The main purpose of Lemma 2 is a cascade transform of equation \eqref{eq17}, obtained at the first step, into regression equations \eqref{eq18} and \eqref{eq19} at the second step. Based on the obtained regression equations \eqref{eq18} and \eqref{eq19}, the identification laws are introduced:
\begin{equation}\label{eq21}
\begin{gathered}
  {{\dot {\hat \Theta} }_{AB}}\!\left( t \right)\! =\!  - {\gamma _\Theta }\left( t \right){\mathcal{M}_{AB}}\left( t \right)\!\left[ {{\mathcal{M}_{AB}}\!\left( t \right){{\hat \Theta }_{AB}}\left( t \right) \!-\! {\mathcal{Y}_{AB}}\!\left( t \right)} \right]{\text{,}} \\ 
  \dot {\hat L}\left( t \right) =  - {\gamma _L}\left( t \right){\mathcal{M}_L}\left( t \right)\left[ {{\mathcal{M}_L}\left( t \right)\hat L\left( t \right) - {\mathcal{Y}_L}\left( t \right)} \right]{\text{,}} \\ 
  {\gamma _\Theta }\left( t \right){\text{:}} = \left\{ \begin{gathered}
  0,{\text{ if }}\Delta \left( t \right) < \rho  \in \left[ {{\Delta _{{\text{min}}}}{\text{; }}{\Delta _{{\text{max}}}}} \right]{\text{,}} \hfill \\
  \frac{{{\gamma _1} + {\gamma _0}{\lambda _{{\text{max}}}}\left( {\Phi(\hat x,\; u) {\Phi(\hat x,\; u) ^{\text{T}}}} \right)}}{{\mathcal{M}_{AB}^2\left( t \right)}}{\text{ otherwise}}{\text{,}} \hfill \\ 
\end{gathered}  \right.\\
{\gamma _L}\left( t \right){\text{:}} = \left\{ \begin{gathered}
  0,{\text{ if }}\Delta \left( t \right) < \rho  \in \left[ {{\Delta _{{\text{min}}}}{\text{; }}{\Delta _{{\text{max}}}}} \right]{\text{,}} \hfill \\
  \frac{{{\gamma _1} + {\gamma _0}{{\tilde y}^2}}}{{\mathcal{M}_L^2\left( t \right)}}{\text{ otherwise}}{\text{,}} \hfill \\ 
\end{gathered}  \right. \\ 
\end{gathered}   
\end{equation}
where ${\gamma _1} > {\text{0}}{\text{, }}{\gamma _0} > {\text{0}}$ are adaptive gains.

The properties of the observation $\tilde x\left( t \right)$ and parametric ${\tilde \Theta _{AB}}\left( t \right)$ errors are studied in the following theorem.
\begin{theorem}\label{theorem1}
When $\varphi \left( t \right) \in {\rm{FE}}$, the identification laws \eqref{eq21} ensure that the goal \eqref{eq7} is achieved.

Proof of Theorem 1 is given in Appendix.
\end{theorem}

Thus, the proposed adaptive observer is designed under Assumptions 1-3, identifiability criterion \eqref{eq5}, \eqref{eq9}, Hypotheses 1-3  and consists of differential equation \eqref{eq6}, parametrizations \eqref{eq14}-\eqref{eq16}, \eqref{eq17}-\eqref{eq20} and identification laws \eqref{eq21}. When the regressor finite excitation requirement \eqref{eq2} is met, the goal of exponentially stable state observation \eqref{eq7} for a class of linear systems parameterized
by unknown physical parameters is satisfied.

\begin{remark} 
It should be specially noted that proposed observer excludes overparameterization in the sense that identification laws \eqref{eq21} are obtained without direct identification of extended vectors of such intermediate unknown parameters as $\theta $ or ${\psi _{ab}}\left( \theta  \right){\text{, }}\eta (\theta) $.
\end{remark}
\begin{remark} 
It is important to emphasize that any other extension scheme from \cite{b14, b27}, which ensures required boundedness $\Delta \left( t \right) \ge \linebreak \ge {\Delta _{{\rm{min}}}} > 0$ for all $t \ge {t_e}$ under FE condition, can be used instead of \eqref{eq15}. Moreover, for applications, in which PE condition is satisfied, extension scheme from \cite{b7} can also be adopted. So this is another one degree of freedom of the proposed observer.
\end{remark}

\section{Numerical Simulation}
The following observable second-order system has been considered:
\begin{equation}\label{eq22}
\begin{array}{l}
\dot x = {\begin{bmatrix}
0&{{\theta _1} + {\theta _2}}&0\\
{ - {\theta _2}}&0&{{\theta _2}}\\
0&{ - {\theta _3}}&0
\end{bmatrix}} x + {\begin{bmatrix}
0\\
0\\
{{\theta _3}}
\end{bmatrix}}u
 = {\Phi ^{\rm{T}}}{\begin{bmatrix}
{{\theta _1} + {\theta _2}}\\
{{\theta _2}}\\
{{\theta _3}}
\end{bmatrix}}{\rm{,}}\\
y = {\begin{bmatrix}
0&0&1
\end{bmatrix}}x{\rm{,}}
\end{array}
\end{equation}
with its representation in the observer canonical form \eqref{eq8}
\begin{gather*}
\begin{array}{l}
\dot \xi  = A_{0}\xi  +  {\begin{bmatrix}
 0\\
{- \left( {{\theta _1} + {\theta _2} + {\theta _3}} \right){\theta _2}}\\
0
\end{bmatrix}y} + {\begin{bmatrix}
{{\theta _3}}\\
0\\
{{\theta _2}{\theta _3}\left( {{\theta _1} + {\theta _2}} \right)}
\end{bmatrix}} u{\rm{,}}\\
y = {\begin{bmatrix}
1&0&0
\end{bmatrix}}\xi {\rm{,}}
\end{array}
\end{gather*}
where
\begin{gather*}
    \begin{array}{c}
\Phi^{\rm{T}} \left( {x{\rm{,\;}}u} \right) = diag\left\{ {{x_2}{\rm{,\;}}{x_3} - {x_1}{\rm{,\;}}u - {x_2}} \right\}{\rm{,}}\\
{\psi _{ab}}\left( \theta  \right) = col\left\{ { - \left( {{\theta _1} + {\theta _2} + {\theta _3}} \right){\theta _2}{\rm{,\;}}{\theta _3}{\rm{,\;}}{\theta _3}{\theta _2}\left( {{\theta _2} + {\theta _1}} \right)} \right\}.
\end{array}
\end{gather*}

The conditions \eqref{eq11}, \eqref{eq12} were satisfied for the system under consideration, and consequently, the mappings ${\cal S}\left( {{\psi _{ab}}} \right){\rm{,\;}}{\cal G}\left( {{\psi _{ab}}} \right){\rm{,\;}}{\Pi _\theta }\left( \Delta  \right)$ existed and were defined as follows:
    \begin{equation}\label{eq23}
\begin{array}{c}
{\cal S}\left( {{\psi _{ab}}} \right) = {\begin{bmatrix}
{{\psi _{2ab}}{{\left( {{\psi _{1ab}}{\psi _{2ab}} + {\psi _{3ab}}} \right)}^2} - \psi _{2ab}^4{\psi _{3ab}}}\\
{ - {\psi _{1ab}}{\psi _{2ab}} - {\psi _{3ab}}}\\
{{\psi _{2ab}}{\psi _{1ab}}}
\end{bmatrix}} {\rm{, }}\\
{\Pi _\theta }\left( \Delta  \right) = diag\left\{ {{\Delta ^5}{\rm{,\;}}{\Delta ^2}{\rm{,\;}}{\Delta ^2}} \right\},\\
{\cal G}\left( {{\psi _{ab}}} \right) = diag\left\{ {\psi _{2ab}^3\left( {{\psi _{1ab}}{\psi _{2ab}} + {\psi _{3ab}}} \right){\rm{,}}\psi _{2ab}^2{\rm{,}}{\psi _{1ab}}} \right\}.
\end{array}
\end{equation}

In their turn, the mappings ${{\cal T}_{\cal S}}\left( . \right){\rm{,\;}}{{\cal T}_{\cal G}}\left( . \right){\rm{,\;}}{{\cal T}_\Theta }\left( . \right)$ were defined as:
   \begin{equation}\label{eq24}
\begin{array}{c}
{{\cal T}_{\cal S}}\left( {{{\overline \Xi }_{\cal S}}\left( \Delta  \right){\cal Y}} \right) =  {\begin{bmatrix}
{{{\cal Y}_2}{{\left( {{{\cal Y}_1}{{\cal Y}_2} + \Delta {{\cal Y}_3}} \right)}^2} - {\cal Y}_2^4{{\cal Y}_3}}\\
{ - {{\cal Y}_1}{{\cal Y}_2} - \Delta {{\cal Y}_3}}\\
{{{\cal Y}_2}{{\cal Y}_1}}
\end{bmatrix}}{\rm{,}}\\
{{\cal T}_{\cal G}}\left( {{{\overline \Xi }_{\cal G}}\left( \Delta  \right){\cal Y}} \right) = { {\begin{bmatrix}
{{\cal Y}_2^3\left( {{{\cal Y}_1}{{\cal Y}_2} + \Delta {{\cal Y}_3}} \right)}&{{\cal Y}_2^2}&{\Delta {{\cal Y}_1}}
\end{bmatrix}}^{\rm{T}}}{\rm{,  }}\\
{{\cal T}_\Theta }\left( {{{\overline \Xi} _\Theta }\left( {\cal{M_{\theta}}}  \right){{\cal Y}_\theta }} \right) = { {\begin{bmatrix}
{{{\cal Y}_{1\theta }} + {{\cal Y}_{2\theta }}}&{{{\cal Y}_{2\theta }}}&{{{\cal Y}_{3\theta }}}
\end{bmatrix}}^{\rm{T}}},
\end{array}
\end{equation}
{where
\begin{gather*}
    {{{\overline \Xi }_{\cal S}}\left( \Delta  \right){\cal Y}}=\begin{bmatrix}
        {\cal Y}_{1}&{\cal Y}_{2}&{\cal Y}_{3}&{\Delta}{\cal Y}_{3}
    \end{bmatrix}^{\rm{T}},\\
    {{{\overline \Xi }_{\cal G}}\left( \Delta  \right){\cal Y}} =\begin{bmatrix}
        {\cal Y}_{1}&{\cal Y}_{2}&{\Delta}{\cal Y}_{1}&{\Delta}{\cal Y}_{3}
    \end{bmatrix}^{\rm{T}},\\
    {{{\overline \Xi} _\Theta }\left( {\cal{M_{\theta}}}  \right){{\cal Y}_\theta }} =\begin{bmatrix}
        {\cal Y}_{1\theta}&{\cal Y}_{2\theta}&{\cal Y}_{3\theta}
    \end{bmatrix}^{\rm{T}}.
\end{gather*}}

The mapping $L(\theta)$ for the system \eqref{eq22} was defined using the following pole placement equation
\begin{equation*}
    \det \left\{ {s{I_n} - A\left( \theta  \right) + L\left( \theta  \right){C^{\rm{T}}}} \right\} = \det \left\{ {s{I_n} - {A_{ref}}} \right\}=(s+a_{m})^3.
\end{equation*}

Consequently, the mappings ${\cal Q}\left( \theta  \right){\rm{,\;}}{\cal P}\left( \theta  \right)$ from Hypothesis 3 took the following form:
\begin{equation}\label{eq25}
\begin{array}{c}
{\cal P}\left( \theta  \right) = diag\left\{ {{\theta _2}{\theta _3}{\rm{,\;}}{\theta _3}{\rm{,\;}}{\theta _1}} \right\}{\rm{,}}\\
{\cal Q}\left( \theta  \right) = \left[ {\begin{array}{*{20}{c}}
{a_m^3 - \left( {3{a_m}{\theta _2} + 3{\theta _1}} \right){\theta _2}}\\
{ - 3a_m^2 + \left( {{\theta _1} + {\theta _2} + {\theta _3}} \right){\theta _2}}\\
{3{a_m}{\theta _1}}
\end{array}} \right]{\rm{, }}
\end{array}
\end{equation}
and the mappings ${{\cal T}_{\cal P}}\left( . \right){\rm{,\;}}{{\cal T}_{\cal Q}}\left( . \right)$, consequently, were defined as:
\begin{equation}\label{eq26}
\begin{array}{c}
{{\cal T}_{\cal P}}\left( {{{\overline \Xi }_{\cal P}}\left( {{{\cal M}_\theta }} \right){{\cal Y}_\theta }} \right) = {{\begin{bmatrix}
{{{\cal Y}_{2\theta }}{{\cal Y}_{3\theta }}}&{{{\cal M}_\theta }{{\cal Y}_{3\theta }}}&{{{\cal Y}_{1\theta }}}
\end{bmatrix}}^{\rm{T}}}{\rm{,}}\\
{\rm{ }}{{\cal T}_{\cal Q}}\left( {{{\overline \Xi }_{\cal Q}}\left( {{{\cal M}_\theta }} \right){{\cal Y}_\theta }} \right) = \\
 =  {\begin{bmatrix}
{a_m^3{\cal M}_\theta ^2 - \left( {3{a_m}{{\cal Y}_{2\theta }} + 3{{\cal Y}_{1\theta }}} \right){{\cal Y}_{2\theta }}}\\
{ - 3a_m^2{\cal M}_\theta ^2 + \left( {{{\cal Y}_{1\theta }} + {{\cal Y}_{2\theta }} + {{\cal Y}_{3\theta }}} \right){{\cal Y}_{2\theta }}}\\
{3{a_m}{{\cal Y}_{1\theta }}}
\end{bmatrix}},
\end{array}
\end{equation}
{where
\begin{gather*}
     {{{\overline \Xi }_{\cal P}}\left( {{{\cal M}_\theta }}\right){{\cal Y}_\theta }}=
\begin{bmatrix}
  {{\cal Y}_{1\theta} }&{{\cal Y}_{2\theta}}&{{\cal Y}_{3\theta}}&{{{\cal M}_\theta }{{\cal Y}_{3\theta }}} 
\end{bmatrix}^{\rm{T}},\\
{{{\overline \Xi }_{\cal Q}}\left( {{{\cal M}_\theta }} \right){{\cal Y}_\theta }}=
\begin{bmatrix}
{{\cal Y}_{1\theta} }&{{\cal Y}_{2\theta}}&{{\cal Y}_{3\theta}}
\end{bmatrix}^{\rm{T}}.
\end{gather*}
}
The control law was chosen to be proportional $u =  - 25\left( {r - y} \right)$. The reference signal $r$, parameters of the system \eqref{eq22} and mappings \eqref{eq25}, \eqref{eq26} were picked as:
\begin{equation}\label{eq27}
\begin{array}{c}
r = 10^2 + 2.5{e^{ - t}}{\rm{sin}}\left({10t}\right){\rm{,}\;}{\theta _1}\! = \!{a_m} = {\theta _2} = 1,{\rm{}}{\theta _3} =  - 1,\\
{x_0} = {{\begin{bmatrix}
{ - 1}&0&2
\end{bmatrix}}^{\rm{T}}}.
\end{array}
\end{equation}

The parameters of the filters \eqref{eq15}, \eqref{eq16} and identification laws \eqref{eq21} were set as follows:
\begin{equation}\label{eq28}
K = {{\begin{bmatrix}
3&3&1
\end{bmatrix}}^{\rm{T}}}{\rm{,}}\begin{array}{*{20}{c}}
{k = {{10}^7}}\\
{\sigma  = 5}
\end{array}{\rm{,\;}}\rho  = {10^{ - 1}}{\rm{,\;}}\begin{array}{*{20}{c}}
{{\gamma _1} = 1}\\
{{\gamma _0} = {{10}^{ - 4}}}
\end{array}.
\end{equation}

Figures 1 and 2 depict the transient curves of the observation error $\tilde x\left( t \right)$ and estimates $\hat \Theta_{AB} \left( t \right){\rm{,\;}}\hat L\left( t \right)$.
\begin{figure}[!thpb]\label{figure1}
\begin{center}
\includegraphics[scale=0.6]{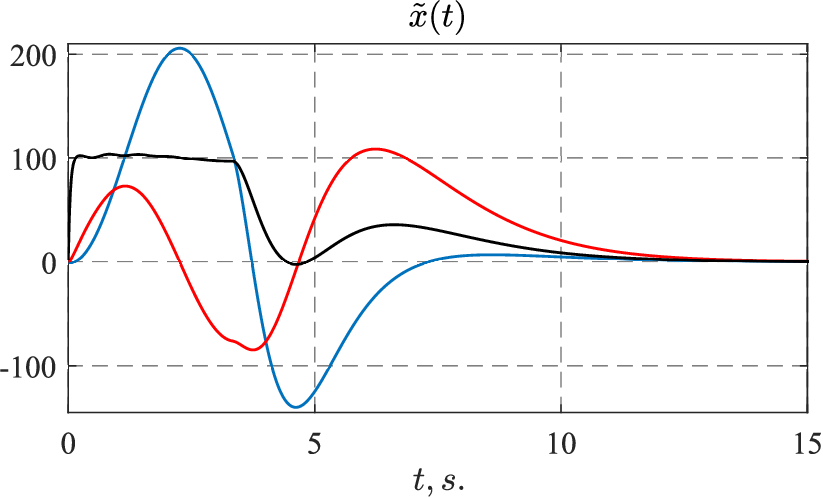}
\caption{{Transient curve of the observation error $\tilde x\left( t \right)$}} 
\end{center}
\end{figure}

\begin{figure}[!thpb]\label{figure2}
\begin{center}
\includegraphics[scale=0.6]{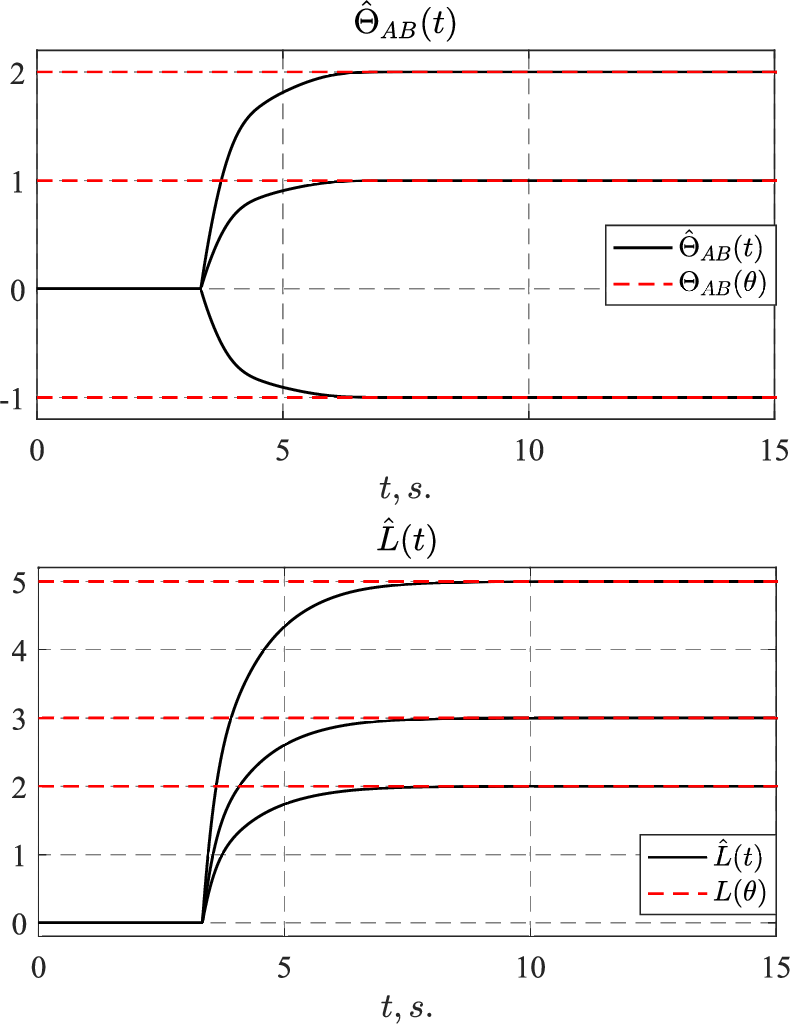}
\caption{{Transient curves of the estimates $\hat \Theta_{AB} \left( t \right){\rm{,\;}}\hat L\left( t \right)$.}} 
\end{center}
\end{figure}

The obtained results corroborated the conclusions made in Theorem 1. After the condition \eqref{eq2} had been met over the time range $\left[ {0{\rm{; 3}}} \right]$, the proposed adaptive observer \eqref{eq6} ensured the goal \eqref{eq7} achievement.

{In order to compare the proposed adaptive observer \eqref{eq6} + \eqref{eq21} with the one with dynamic inversion-based change of coordinates, it is interesting to see what conditions are required for safe implementation of substitutions (10b) for the example under consideration. Having} {analyzed the decomposition  \eqref{eq12}, \eqref{eq23}, to avoid singularity in the function ${\cal{F}}\left(\hat{\psi}_{ab}\right)$, for all $t\ge t_{0}$ the estimates $\hat{\psi}_{ab}$ need to satisfy:}
\begin{gather*}{
    \begin{array}{c}
{\rm{sign}}\left( {{\psi _{1ab}}} \right) = {\rm{sign}}\left( {{{\hat \psi }_{1ab}}} \right){\rm{,\;sign}}\left( {{\psi _{2ab}}} \right) = {\rm{sign}}\left( {{{\hat \psi }_{2ab}}} \right),\\
{\rm{sign}}\left( {{\psi _{1ab}}{\psi _{2ab}} + {\psi _{3ab}}} \right) = {\rm{sign}}\left( {{{\hat \psi }_{1ab}}{{\hat \psi }_{2ab}} + {{\hat \psi }_{3ab}}} \right).
\end{array}}
\end{gather*}

{In addition to this, to avoid singularity in the function $L\left({\cal{F}}\left(\hat{\psi}_{ab}\right)\right)$, for all $t\ge t_{0}$ the estimates $\hat{\psi}_{ab}$ need to satisfy ${\rm{sign}}\left(\theta\right)={\rm{sign}}\left({\cal{F}}\left(\hat{\psi}_{ab}\right)\right)$. For existing identification laws \cite{b2, b3, b4, b5, b6, b7, b8, b9, b10, b11, b14, b27, b29} these strict conditions are almost always violated. Under condition that Hypotheses 1-3 are satisfied, the proposed solution \eqref{eq6} + \eqref{eq21} are fully free from this drawback.}

\section{Conclusion and Future Work}
To reconstruct the unobservable original states of completely observable linear time-invariant systems parameterized by unknown physical parameters, a method to design adaptive observers is proposed in this note. Unlike existing adaptive solutions, the system state-space matrices {\it{A, B}} are not restricted to be represented in the observer canonical form to implement the observer. The approach is applicable to above-mentioned systems if: ({\it{i}}) the condition \eqref{eq5}, \eqref{eq9} of identifiability of the physical parameters $\theta$ is satisfied, ({\it{ii}}) the Luenberger correction gain, the system matrices and the numerator/denominator polynomials of the system are dependent on the physical parameters in polynomial manner (Hypotheses 1-3). If the sufficiently weak finite excitation condition \eqref{eq2} is met, the proposed observer ensures exponential convergence of the observation error to zero. 

In the further research it is planned to extend the obtained results to the cases of external perturbations and systems with time-varying parameters. 
\appendices

\renewcommand{\theequation}{A\arabic{equation}}
\setcounter{equation}{0}  

\section*{Appendix}
{\it{Proof of Lemma 1.}} Let the following signal be introduced:
    \begin{equation}\label{eqA1}
\tilde \xi \left( t \right) = \xi \left( t \right) - z\left( t \right) - \Omega \left( t \right){\psi _a}\left( \theta  \right) - P\left( t \right){\psi _b}\left( \theta  \right).
\end{equation}

Equation \eqref{eqA1} is differentiated with respect to time:
    \begin{equation}\label{eqA2}
\begin{array}{l}
{\dot {\tilde \xi}} \left( t \right) = {A_0}\xi \left( t \right) + {\psi _a}\left( \theta  \right)y\left( t \right) + {\psi _b}\left( \theta  \right)u\left( t \right) - \\
 - {A_K}z\left( t \right) - Ky\left( t \right) - \left( {{A_K}\Omega \left( t \right) + {I_n}y\left( t \right)} \right){\psi _a}\left( \theta  \right) - \\
 - \left( {{A_K}P\left( t \right) + {I_n}u\left( t \right)} \right){\psi _b}\left( \theta  \right) = \\
 = {A_0}\xi \left( t \right) - {A_K}z\left( t \right) - Ky\left( t \right) - {A_K}\Omega \left( t \right){\psi _a}\left( \theta  \right) - \\
 - {A_K}P\left( t \right){\psi _b}\left( \theta  \right) = {A_K}\tilde \xi \left( t \right).
\end{array}
\end{equation}

The solution of equation \eqref{eqA2} is substituted into \eqref{eqA1}, and the obtained result is multiplied by $C_0^{\rm{T}}$, from which we have a measurable regression equation:
    \begin{equation}\label{eqA3}
q\left( t \right) = {\varphi ^{\rm{T}}}\left( t \right)\eta(\theta).
\end{equation}

Having applied the filtering \eqref{eq16} to it, the extended regression equation is obtained:
\begin{equation}\label{eqA4}
\overline q\left( t \right) = \overline \varphi \left( t \right)\eta(\theta).
\end{equation}

Equation \eqref{eqA4} is then multiplied by $k{{\cal L}_{ab}}{{\cal L}_0}{\rm{adj}}\left\{ {\overline \varphi \left( t \right)} \right\}$ to obtain the regression equation \eqref{eq14}. The fact that for all $t \ge {t_e}$ the condition $\Delta \left( t \right) \ge {\Delta _{{\rm{min}}}} > 0$  is satisfied in case $\varphi \left( t \right) \in {\rm{FE}}$ has been proved, for an instance, in Proposition 4 in \cite{b28}.

{\it{Proof of Lemma 2.}} In accordance with Definition 1 and Hypothesis 2 and owing to:
  \begin{equation}\label{eqA5}
\begin{array}{c}
{\Xi _{\cal S}}\left( {\Delta } \right) = {{\overline \Xi }_{\cal S}}\left( {\Delta } \right)\Delta {\rm{,  }}\\{\Xi _{\cal G}}\left( {\Delta } \right) = {{\overline \Xi }_{\cal G}}\left( {\Delta } \right)\Delta {\rm{,  }}\\
{\cal Y}\left( t \right) = \Delta \left( t \right){\psi _{ab}}\left( \theta  \right)
\end{array}
\end{equation}
the following equality is obtained from \eqref{eq12}:
  \begin{equation}\label{eqA6}
{{\cal T}_{\cal S}}\left( {{{\overline \Xi }_{\cal S}}\left( {\Delta } \right){\cal Y}} \right) = {{\cal T}_{\cal G}}\left( {{{\overline \Xi }_{\cal G}}\left( {\Delta } \right){\cal Y}} \right)\theta.
\end{equation}

Then, having multiplied \eqref{eqA6} by ${\rm{adj}}\left\{ {{{\cal T}_{\cal G}}\left( {{{\overline \Xi }_{\cal G}}\left( {\Delta } \right){\cal Y}} \right)} \right\}$, the regression equation \eqref{eq17} is obtained, from which, together with \eqref{eq11}, we have:
  \begin{equation}\label{eqA7}
{{\cal T}_\Theta }\left( {{{\overline \Xi }_\Theta }\left( {{{\cal M}_\theta }} \right){{\cal Y}_\theta }} \right) = {\Pi _\Theta }\left( {{{\cal M}_\theta }} \right){\Theta _{AB}}\left( \theta  \right).
\end{equation}

Equation \eqref{eqA7} is multiplied by ${\rm{adj}}\left\{ {{\Pi _\Theta }\left( {{{\cal M}_\theta }} \right)} \right\}$ to obtain the regression equation \eqref{eq18}. Following Definition 1 and Hypothesis 3, similar to \eqref{eqA6} and \eqref{eqA7}, it is written:
  \begin{equation}\label{eqA8}
{{\cal T}_{\cal Q}}\left( {{{\overline \Xi }_{\cal Q}}\left( {{{\cal M}_\theta }} \right){{\cal Y}_\theta }} \right) = {{\cal T}_{\cal P}}\left( {{{\overline \Xi }_{\cal P}}\left( {{{\cal M}_\theta }} \right){{\cal Y}_\theta }} \right)L(\theta).
\end{equation}

Having multiplied \eqref{eqA8} by ${\rm{adj}}\left\{ {{{\cal T}_{\cal P}}\left( {{{\overline \Xi }_{\cal P}}\left( {{{\cal M}_\theta }} \right){{\cal Y}_\theta }} \right)} \right\}$, the regression equation \eqref{eq19} is obtained. Following the proof of Lemma 1, when $\varphi \left( t \right) \in {\rm{FE}}$,  $\Delta \left( t \right) \ge {\Delta _{{\rm{min}}}} > 0{\rm{\;}}\forall t \ge {t_e}{\rm{ }}$ holds, and by Hypotheses 1-3:
\begin{gather*}
\begin{array}{c}
{\rm{det}}^2\!\left\{ {{\cal G}\left( {{\psi _{ab}}} \right)} \right\}\! > \!0,{\rm{ det}}^2\!\left\{ {{\cal P}\left( \theta  \right)} \right\} > 0,{\rm{ det}}\left\{ {{\Pi _\theta }\left( {\Delta } \right)} \right\} \!\ge\! {\Delta ^{{\ell _\theta }}}\left( t \right){\rm{, }}\\
{\rm{det}}\left\{ {{\Pi _\Theta }\left( {{{\cal M}_\theta }} \right)} \right\} \ge {\cal M}_\theta ^{{\ell _\Theta }}\left( t \right){\rm{, det}}\left\{ {{\Pi _L}\left( {{{\cal M}_\theta }} \right)} \right\} \ge {\cal M}_\theta ^{{\ell _L}}\left( t \right),
\end{array}
\end{gather*}
from which $\forall t \ge {t_e}$ the following inequalities hold in case \linebreak $\varphi \left( t \right) \in {\rm{FE}}$:
\begin{equation}\label{eqA9}
\begin{array}{c}
\left| {{{\cal M}_\theta }\left( t \right)} \right| = \left| {{\rm{det}}\left\{ {{{\cal T}_{\cal G}}\left( {{\Xi _{\cal G}}\left( {\Delta } \right){\psi _{ab}}} \right)} \right\}} \right| = \\
 = \left| {{\rm{det}}\left\{ {{\Pi _\theta }\left( \Delta  \right)} \right\}{\rm{det}}\left\{ {{\cal G}\left( {{\psi _{ab}}} \right)} \right\}} \right| \ge \left| {{\rm{det}}\left\{ {{\cal G}\left( {{\psi _{ab}}} \right)} \right\}} \right|\Delta _{\min }^{{\ell _\theta }} > 0,\\
\left| {{{\cal M}_{AB}}\left( t \right)} \right| = \left| {{\rm{det}}\left\{ {{\Pi _\Theta }\left( {{{\cal M}_\theta }} \right)} \right\}} \right| \ge \left| {{\cal M}_\theta ^{{\ell _\Theta }}\left( t \right)} \right| \ge \\
 \ge \left| {{\rm{de}}{{\rm{t}}^{{\ell _\Theta }}}\left\{ {{\cal G}\left( {{\psi _{ab}}} \right)} \right\}} \right|\Delta _{\min }^{{\ell _\theta }{\ell _\Theta }} = \underline {{{\cal M}_{AB}}}  > 0,\\
\left| {{{\cal M}_L}\left( t \right)} \right| = \left| {{\rm{det}}\left\{ {{{\cal T}_{\cal P}}\left( {{\Xi _{\cal P}}\left( {{{\cal M}_\theta }} \right)\theta } \right)} \right\}} \right| = \\
 = \left| {{\rm{det}}\left\{ {{\cal P}\left( \theta  \right)} \right\}{\rm{det}}\left\{ {{\Pi _L}\left( {{{\cal M}_\theta }} \right)} \right\}} \right| \ge \left| {{\rm{det}}\left\{ {{\cal P}\left( \theta  \right)} \right\}} \right|\left| {{\cal M}_\theta ^{{\ell _L}}\left( t \right)} \right| \ge \\
 \ge \left| {{\rm{det}}\left\{ {{\cal P}\left( \theta  \right)} \right\}} \right|\left| {{\rm{de}}{{\rm{t}}^{{\ell _L}}}\left\{ {{\cal G}\left( {{\psi _{ab}}} \right)} \right\}} \right|\Delta _{\min }^{{\ell _\theta }{\ell _L}} = \underline {{{\cal M}_L}}  > 0,
\end{array}
\end{equation}
which completes the proof of Lemma 2.

{\it{Proof of Theorem 1.}} 
To prove the theorem, first of all, the error equation between $\hat x(t)$ and $x(t)$ is written as:
\begin{equation}\label{eqA10}
\begin{gathered}
  {\dot {\tilde x}}\left( t \right) = {\Phi ^{\text{T}}}\left( {\hat x{\text{, }}u} \right){{\hat \Theta }_{AB}}\left( t \right) - \hat L\left( t \right)\tilde y\left( t \right) - {\Phi ^{\text{T}}}\left( {x{\text{, }}u} \right){\Theta _{AB}}(\theta)\\
   = {\Phi ^{\text{T}}}\left( {\hat x{\text{, }}u} \right){{\hat \Theta }_{AB}}\left( t \right) - \hat L\left( t \right)\tilde y\left( t \right) - {\Phi ^{\text{T}}}\left( {x{\text{, }}u} \right){\Theta _{AB}}(\theta)\pm \\
   \pm {\Phi ^{\text{T}}}\left( {\hat x{\text{, }}u} \right){\Theta _{AB}}(\theta) =  {\Phi ^{\text{T}}}\left( {\hat x{\text{, }}u} \right){{\tilde \Theta }_{AB}}\left( t \right) - \hat L\left( t \right)\tilde y\left( t \right)- \\ - {\Phi ^{\text{T}}}\left( {x{\text{, }}u} \right){\Theta _{AB}}(\theta) + {\Phi ^{\text{T}}}\left( {\hat x{\text{, }}u} \right){\Theta _{AB}}(\theta) =  \\
   = A\left( \theta  \right)\tilde x\left( t \right) + {\Phi ^{\text{T}}}\left( {\hat x{\text{, }}u} \right){{\tilde \Theta }_{AB}}\left( t \right) - \hat L\left( t \right)\tilde y\left( t \right) \pm L(\theta)\tilde y\left( t \right)\\
   = {A_m}\tilde x\left( t \right) + {\Phi ^{\text{T}}}\left( {\hat x{\text{, }}u} \right){{\tilde \Theta }_{AB}}\left( t \right) - \tilde L\left( t \right)\tilde y\left( t \right){\text{,}}  \\ 
\end{gathered}
\end{equation}

Then the following quadratic form is introduced:
\begin{equation}\label{eqA11}
\begin{array}{c}
V = {\zeta ^{\rm{T}}}H\zeta  = {\gamma _0}{{\tilde x}^{\rm{T}}}P\tilde x + {{\tilde \Theta }^{\rm{T}}}\tilde \Theta  + {{\tilde L}^{\rm{T}}}\tilde L{\rm{, }}\\
H = {\rm{blockdiag}}\left\{ {{\gamma _0}P{\rm{,\;}}{I_n}{\rm{,\;}}{I_n}} \right\}{\rm{,}}
\end{array}
\end{equation}
where $ \zeta \left( t \right) = {{\begin{bmatrix}
{{{\tilde x}^{\rm{T}}}\left( t \right)}&{{{\tilde \Theta }^{\rm{T}}}\left( t \right)}&{{{\tilde L}^{\rm{T}}}\left( t \right)}
\end{bmatrix}}^{\rm{T}}}$ and $P$ is a solution of the following set of equations, taking into consideration that we have chosen $K = {k^2}{I_{n \times n}}{\rm{,\;}}D = 0.5{k^2}{I_{n \times n}}{\rm{,\;}}k = 1,{\rm{\;}}B = {I_{n \times n}}$:
\begin{gather*}
    \begin{array}{c}
A_m^{\rm{T}}P + P{A_m} =  - Q{Q^{\rm{T}}} - \mu P{\rm{,\;}}P{I_{{\mathop{n}\nolimits}  \times n}} = QK{\rm{,\;}}\\
{K^{\rm{T}}}K = D + {D^{\rm{T}}},
\end{array}
\end{gather*}
which is equivalent to the Riccati equation with bias $A_m^{\rm{T}}P + P{A_m} + \\ + P{P^{\rm{T}}} + \mu P = {0_{n \times n}}.$

The derivative of \eqref{eqA11} with respect to \eqref{eqA10} is written as:
\begin{equation}\label{eqA12}
\begin{array}{c}
\dot V = {\gamma _0}\left[ {{{\tilde x}^{\rm{T}}}\left( {A_m^{\rm{T}}P + P{A_m}} \right)\tilde x + {{\tilde x}^{\rm{T}}}P{\Phi ^{\rm{T}}}{{\tilde \Theta }_{AB}}} \right. + \\
 + \left. {\tilde \Theta _{AB}^{\rm{T}}\Phi P\tilde x - {{\tilde L}^{\rm{T}}}\tilde yP\tilde x - {{\tilde x}^{\rm{T}}}P\tilde L\tilde y} \right] + 2\tilde \Theta _{AB}^{\rm{T}}{{{\dot {\tilde{ \Theta}}}}_{AB}} + \\
 + 2{{\tilde L}^{\rm{T}}}{\dot{ \tilde L}} = {\gamma _0}\left[ { - \mu {{\tilde x}^{\rm{T}}}P\tilde x - {{\tilde x}^{\rm{T}}}Q{Q^{\rm{T}}}\tilde x + } \right.\\
\left. {2{{\tilde x}^{\rm{T}}}P{\Phi ^{\rm{T}}}{{\tilde \Theta }_{AB}} - 2{{\tilde x}^{\rm{T}}}Q\tilde L\tilde y} \right] + 2\tilde \Theta _{AB}^{\rm{T}}{{{\dot{ \tilde \Theta}} }_{AB}} + 2{{\tilde L}^{\rm{T}}}{\dot {\tilde L}} = \\
 = {\gamma _0}\left[ { - \mu {{\tilde x}^{\rm{T}}}P\tilde x - {\textstyle{1 \over 2}}{{\tilde x}^{\rm{T}}}Q{Q^{\rm{T}}}\tilde x + 2{{\tilde x}^{\rm{T}}}Q{\Phi ^{\rm{T}}}{{\tilde \Theta }_{AB}} - } \right.\\
\left. { - {\textstyle{1 \over 2}}{{\tilde x}^{\rm{T}}}Q{Q^{\rm{T}}}\tilde x - 2{{\tilde x}^{\rm{T}}}Q\tilde L\tilde y \pm 2{{\tilde L}^{\rm{T}}}{{\tilde y}^2}\tilde L } \right] + 2{{\tilde L}^{\rm{T}}}{\dot{ \tilde L}} \pm \\
 \pm 2{\gamma _0}\tilde \Theta _{AB}^{\rm{T}}\Phi {\Phi ^{\rm{T}}}{{\tilde \Theta }_{AB}} + 2\tilde \Theta _{AB}^{\rm{T}}{{{\dot {\tilde \Theta}} }_{AB}}.
\end{array}
\end{equation}
Completing the square in \eqref{eqA12}, it is obtained:
\begin{equation}\label{eqA13}
\begin{array}{c}
\dot V = 2\tilde \Theta _{AB}^{\rm{T}}{{{\dot{ \tilde \Theta}}}_{AB}} + 2{{\tilde L}^{\rm{T}}}{\dot {\tilde L}} + {\gamma _0}\left[ { - \mu {{\tilde x}^{\rm{T}}}P\tilde x} \right. - \\
{\left( {{\textstyle{1 \over {\sqrt 2 }}}{{\tilde x}^{\rm{T}}}Q - \sqrt 2 \tilde \Theta _{AB}^{\rm{T}}\Phi } \right)^2} - {\left( {{\textstyle{1 \over {\sqrt 2 }}}{{\tilde x}^{\rm{T}}}Q + \sqrt 2 {{\tilde L}^{\rm{T}}}\tilde y} \right)^2}\\
\left. { + 2\tilde \Theta _{AB}^{\rm{T}}\Phi {\Phi ^{\rm{T}}}{{\tilde \Theta }_{AB}} + 2{{\tilde L}^{\rm{T}}}{{\tilde y}^2}\tilde L} \right] = \\
 = 2\tilde \Theta _{AB}^{\rm{T}}{{{\dot {\tilde \Theta}} }_{AB}} + 2{{\tilde L}^{\rm{T}}}{\dot {\tilde L}} + {\gamma _0}\left[ { - \mu {{\tilde x}^{\rm{T}}}P\tilde x} \right. + \\
\left. { + 2\tilde \Theta _{AB}^{\rm{T}}{\Phi ^{\rm{T}}}\Phi {{\tilde \Theta }_{AB}} + 2{{\tilde L}^{\rm{T}}}{{\tilde y}^2}\tilde L} \right].
\end{array}
\end{equation}

Following the proof of Lemma 2, in case $ \varphi \left( t \right) \in {\rm{FE}}$ the inequalities \eqref{eqA9} hold, therefore, in accordance with the definition of the adaptive gains, we have from \eqref{eqA13} and \eqref{eq21} for all $t \ge {t_e}$ that:
\begin{gather*}
\begin{array}{c}
\dot V =  - \mu {\gamma _0}{{\tilde x}^{\rm{T}}}P\tilde x + 2{\gamma _0}\tilde \Theta _{AB}^{\rm{T}}\Phi {\Phi ^{\rm{T}}}{{\tilde \Theta }_{AB}} + 2{\gamma _0}{{\tilde L}^{\rm{T}}}{{\tilde y}^2}\tilde L - \\
 - 2\tilde \Theta _{AB}^{\rm{T}}\left( {{\gamma _1} + {\gamma _0}{\lambda _{{\rm{max}}}}\left( {\Phi {\Phi ^{\rm{T}}}} \right)} \right){{\tilde \Theta }_{AB}} - 2{{\tilde L}^{\rm{T}}}\left( {{\gamma _1} + {\gamma _0}{{\tilde y}^2}} \right)\tilde L \le \\
 \le  - {\kappa _{{\rm{min}}}}V{\rm{,\;}}{\kappa _{{\rm{min}}}} = {\rm{min}}\left\{ {{\textstyle{{\mu {\gamma _0}{\lambda _{{\rm{min}}}}\left( P \right)} \over {{\lambda _{{\rm{max}}}}\left( P \right)}}}{\rm{, 2}}{\gamma _1}} \right\}{\rm{,}}
\end{array}    
\end{gather*}
from which it follows that the goal \eqref{eq7} is met.


\begin{thebibliography}{00}
\bibitem{b1} D. Dochain, ``State and parameter estimation in chemical and biochemical processes: a tutorial,'' \emph{Journal of process control}, vol. 13, no. 8, pp. 801--818, 2003.
\bibitem{b2} Carroll R., Lindorff  D., ``An adaptive observer for single-input single-output linear systems,'' \emph{IEEE Transactions on Automatic Control}, vol. 18, no. 5, pp. 428--435, 1973.
\bibitem{b3} Luders G., Narendra K. S., ``An adaptive observer and identifier for a linear system,'' \emph{IEEE Transactions on Automatic Control}, vol. 18, no. 5, pp. 496--499, 1973.
\bibitem{b4} Narendra K. S., Valavani L. S., ``Stable adaptive observers and controllers,'' \emph{Proceedings of IEEE}, vol. 64, no.8, pp. 1198--1208, 1976.
\bibitem{b5} Kudva P., Narendra K. S., ``Synthesis of an adaptive observer using Lyapunov's direct method,'' \emph{International Journal of Control}, vol. 18, no. 6, pp. 1201--1210, 1973.
\bibitem{b6}Narendra K. S., Kudva P., ``Stable adaptive schemes for system identification and control-Part I,'' \emph{IEEE Transactions on Systems, Man, and Cybernetics}, no. 6, pp. 542--551, 1974.
\bibitem{b7}Kreisselmeier G., ``Adaptive observers with exponential rate of convergence,'' \emph{IEEE Transactions on Automatic Control}, vol. 22, no. 1, pp. 2--8, 1977.
\bibitem{b8} A. Katiyar, S. B. Roy, and S. Bhasin, ``Initial Excitation Based Robust Adaptive Observer for MIMO LTI Systems,'' \emph{IEEE Trans. on Automatic Control,} 2022. Early access.
\bibitem{b9} A. Katiyar, S. B. Roy, and S. Bhasin, ``Finite excitation based robust adaptive observer for MIMO LTI systems,'' \emph{Int. J. of Adaptive Control and Signal Proc.,} vol.36, no. 2, pp. 180--197, 2022.
\bibitem{b10} A. Katiyar, S. B. Roy, and S. Bhasin, ``Initial Excitation Based Fast Adaptive Observer,'' in \emph{Proc. Europ. Control Conf.,} London, 2022, pp. 1--8.
\bibitem{b11} A. Katiyar, S. B. Roy, and S. Bhasin, ``Initial excitation based adaptive observer with multiple switching,'' in \emph{Proc. IEEE Conf. on Decision and Control,} Nice, 2019, pp. 2910--2915.
\bibitem{b12} P. Tomei, and R. Marino, ``An enhanced feedback adaptive observer for nonlinear systems with lack of persistency of excitation,'' \emph{IEEE Trans. on Automatic Control,} 2022. Early access.
\bibitem{b13} Q. Zhang, and F. Giri, ``Adaptive Observer with Enhanced Gain to Address Deficient Excitation,'' \emph{IFAC-PapersOnLine,} vol. 55, no. 12, pp. 336--340, 2022.
\bibitem{b14} A. Bobtsov, A. Pyrkin, A. Vedyakov, A. Vediakova, and S. Aranovskiy, ``A Modification of Generalized Parameter-Based Adaptive Observer for Linear Systems with Relaxed Excitation Conditions'' \emph{IFAC-PapersOnLine,} vol. 55, no. 12, pp. 324--329, 2022.
\bibitem{b15} R. Ortega, S. Aranovskiy, A. Pyrkin, A. Astolfi, A. Bobtsov, ``New results on parameter estimation via dynamic regressor extension and mixing: Continuous and discrete-time cases,'' \emph{IEEE Trans. on Automatic Control,} vol.66, no. 5, pp. 2265--2272, 2020.
\bibitem{b16} S. Aranovskiy, A. Bobtsov, R. Ortega, A. Pyrkin, ``Parameters estimation via dynamic regressor extension and mixing,'' in \emph{Proc. American Control Conference ,} Boston, 2016, pp. 6971--6976.
\bibitem{b17} Y. M. Cho, R. Rajamani, `` A systematic approach to adaptive observer synthesis for nonlinear systems,'' \emph{IEEE Trans. on Automatic Control,} vol.42, no. 4, pp. 534--537, 1997.
\bibitem{b18} Cecilia A., Costa-Castelló R., ``Addressing the relative degree restriction in nonlinear adaptive observers: A high-gain observer approach,'' \emph{Journal of the Franklin Institute,} vol.359, no. 8, pp. 3857--3882, 2022.
\bibitem{b19} Farza M., M’Saad M., Maatoug T., Kamoun M. ``Adaptive observers for nonlinearly parameterized class of nonlinear systems,'' \emph{Automatica}, vol. 45, pp. 2292--2299, 2009.
\bibitem{b20} Bin M., Marconi L., ``Model identification and adaptive state observation for a class of nonlinear systems,'' \emph{IEEE Trans. on Automatic Control,} vol. 66, no. 12, pp. 5621--5636, 2020.
\bibitem{b21} Bastin G., Gevers M. R., ``Stable adaptive observers for nonlinear time-varying systems,'' \emph{IEEE Trans. on Automatic Control,} vol. 33, no. 7, pp. 650--658, 1988.
\bibitem{b22} Afri C., Andrieu V., Bako L., Dufour P, `` State and parameter estimation: A nonlinear Luenberger observer approach,'' \emph{IEEE Trans. on Automatic Control,} vol. 62, no. 2, pp. 973--980, 2016.
\bibitem{b23} Grewal M., Glover K., ``Identifiability of linear and nonlinear dynamical systems,'' \emph{IEEE Trans. on Automatic Control,} vol. 21, no. 6, pp. 833--837, 1976.
\bibitem{b24} Lecourtier Y., Walter E., Bertrand P., ``Identifiability testing for state-space models,'' \emph{IFAC Proceedings Volumes,} vol. 15, no. 4, pp. 887--892, 1982.
\bibitem{b25} Lecourtier Y., Waiter E., Raksanyi A., ``Test of Structural Properties of Statespace Models Through Algebraic Computation,'' \emph{IFAC Proceedings Volumes,} vol. 17, no. 2, pp. 545--550, 1984.
\bibitem{b26} A. Glushchenko, V. Petrov, and K. Lastochkin, ``Exponentially stable adaptive control. Part I. Time-invariant plants,'' \emph{Automation and Remote Control,} vol. 83, no. 4, pp. 548--578, 2022.
\bibitem{b29} P. A. Ioannou, and J. Sun, \emph{Robust adaptive control,} Mineola, NY, USA: Courier Corp., 2012.
\bibitem{b27} L. Wang, R. Ortega, A. Bobtsov, J. G. Romero,  B. Yi, ``Identifiability implies robust, globally exponentially convergent on-line parameter estimation: Application to model reference adaptive control,'' \emph{arXiv preprint arXiv:2108.08436} pp.1--16, 2021.
\bibitem{b28} A. Glushchenko, K. Lastochkin, ``Unknown piecewise constant parameters identification with exponential rate of convergence,'' \emph{Int. J. of Adaptive Control and Signal Proc.,} vol.37, no. 1, pp. 315--346, 2023.
\end{thebibliography}
\end{document}